\documentclass[aps,twocolumn]{revtex4}
\usepackage{color}
\usepackage{psfrag}
\usepackage{dcolumn}
\usepackage{bm}
\usepackage{float}
\usepackage[latin1]{inputenc}
\usepackage[spanish,english]{babel}
\usepackage{amsfonts}
\usepackage{amssymb,epsf}
\usepackage{graphicx}
\usepackage{slashed}
\usepackage{epstopdf}
\usepackage{amsmath,amssymb}
\usepackage{pdflscape}
\usepackage{adjustbox}
\usepackage{hyperref}
\usepackage{subcaption}
\usepackage{appendix}
\usepackage{multirow}
\usepackage{xcolor}
\definecolor{darkgreen}{RGB}{0,150,0}
\begin{document}
	
	\title{Microscopic Interaction versus Purely Gravitational Coupling in Strange Quark Stars Admixed with Dark Matter: A One--Fluid and Two--Fluid Comparison}
	
	\author{J. Sedaghat\footnote{
			email address: J.sedaghat@shirazu.ac.ir}, G. H. Bordbar\footnote{
			email address: ghbordbar@shirazu.ac.ir (corresponding author)}, M. Haghighat\footnote{
			email address: m.haghighat@shirazu.ac.ir }, S. M. Zebarjad\footnote{
			email address: zebarjad@shirazu.ac.ir}}

	\affiliation{Physics Department and Biruni Observatory, Shiraz University, Shiraz 71454, Iran}
	\begin{abstract}	
We investigate strange quark stars (SQSs) admixed with scalar dark matter (DM), focusing on the role of microscopic interactions versus purely gravitational coupling. {Specifically, we study whether the model with quark--DM interactions or the non--interacting one is more compatible with current as well as expected observational results from gravitational waves and pulsars.} To this end, two distinct approaches are compared: a one--fluid model, where strange quark matter (SQM) and DM are fully mixed and interact via a Yukawa--type coupling and Bose-Einstein condensation pressure, and a two--fluid model, where SQM and DM possess independent equations of state and couple solely through gravity. For both scenarios, we determine the stellar structural properties and compute the mass--radius ($M-R$) and dimensionless tidal deformability--mass ($\Lambda\text{--}M$) relations, exploring a range of central DM pressure fractions $f_r$ and DM particle masses $m_D$. Our results demonstrate that while the interacting one--fluid model yields viable configurations with $M_{\text{TOV}} > 2 M_\odot$ that satisfy current pulsar $M-R$ measurements alongside the $\Lambda$ constraints from GW170817, the non--interacting two--fluid model more robustly meets both existing GW170817 limits and the tighter $\Lambda$ bounds anticipated from next--generation gravitational wave detectors. {To further demonstrate this, we investigate how both models respond to representative parameter changes, showing that in the single-fluid model, stiffening or softening the equation of state tends to violate either the predicted tighter $\Lambda$ constraint or the $2.0\,M_\odot$ mass limit. In contrast, the two-fluid model can more easily satisfy both requirements.}

\end{abstract}

\maketitle
\section{Introduction}

Dark matter (DM) is one of the most intriguing components of the Universe, accounting for roughly 27\% of its total mass -- energy budget, yet its microscopic nature remains unknown \cite{Bertone2018,Planck2014}. Its presence is inferred from a variety of astrophysical and cosmological observations, including galaxy rotation curves \cite{Zwicky1933,Persic1996}, gravitational lensing \cite{Allen2011}, and the large--scale structure of the cosmos \cite{Frenk2012,Bennett2013}. Among the many candidates proposed, scalar--field dark matter (DM) has emerged as a compelling possibility, motivated both by particle physics extensions beyond the Standard Model (e.g., Higgs--sector extensions) and by its ability to produce macroscopic quantum phenomena such as Bose--Einstein condensation (BEC) in astrophysical environments \cite{Panotopoulos2017,Lopes2018,Matos2000,Matos2001,Bernal2006,Boehm2001,Gottel2024,Branco2012,Gunion1989}.

Compact stars provide a unique natural laboratory for probing the properties of DM under extreme conditions of density and gravity. If DM can accumulate inside such stars, either through capture during their evolution or from formation in a DM--rich environment, it can alter their equilibrium structure, stability, and observable signatures \cite{Chaudhuri2024,Anzuini,Routaray,Das2019,Panotopoulos2017,Karkevandi2022,Karkevandi2024,Karkevandi2024b,Yang2025,sedaghatonefluid}. The degree and nature of these modifications depend on whether DM interacts with the baryonic matter sector only via gravity or also through additional microscopic couplings \cite{Chaudhuri2024,Anzuini,Routaray,Das2019,sedaghatonefluid,Shahrbaf2025}. Therefore,  modelling the DM contribution to the stellar equation of state (EOS) under different interaction assumptions  provides a direct link between theoretical particle physics models and measurable astrophysical properties such as the mass--radius relation and tidal deformability.

Strange quark stars (SQSs) \cite{Michel1988,Drago2001,Kurkela2010,Wang2019,Deb2021,Sedaghat2021,JSedaghat,sedaghat,sedaghatannals} are hypothetical compact stars composed entirely of strange quark matter (SQM) --- a deconfined mixture of $u$, $d$, and $s$ quarks --- which may be the true ground state of strongly interacting matter at high densities \cite{Witten1984,Weber2005}. In the absence of DM, SQSs are described by EOSs derived from phenomenological models such as the NJL model \cite{Mishustin, Hanauske, Buballa2005, Chu2016, ChengMingLi2020, Gholami2025, sedaghatonefluid}, MIT bag model \cite{Hua Li2010, MassIV, Deb2018, Podder2024} or from perturbative QCD (pQCD) calculations \cite{Kurkela2010,Fraga2006,sedaghat,sedaghatannals}. Direct observational confirmation of quark stars remains difficult due to their close resemblance to neutron stars. 
An independent model analysis in Ref.~\cite{Annala} indicates that, provided the conformal bound on the sound speed ($c_s^2 \leq 1/3$) is not strongly violated, massive neutron stars are expected to host sizable quark--matter cores.

Observational constraints from massive pulsars, such as PSR J0740+6620  \cite{Cromartie2019}, PSR~J0030+0451~\cite{Riley2019}, 
PSR~J0437--4715~\cite{PSR J0437 - 4715}, and PSR~J0614--3329~\cite{PSR J0614 - 3329}  along with radius measurements from NICER and the tidal deformability bound from GW170817 \cite{Abbott2017,Abbott2018}, impose stringent requirements on the stiffness of any viable EOS. 
To simultaneously account for these observations, the EOS must be sufficiently stiff to support {such massive compact stars}, while remaining consistent with the relatively small tidal deformability inferred from GW170817. 
EOSs describing SQM in the 
color--flavor--locked (CFL) phase naturally support the existence of massive SQSs \cite{Roupas2021, SedaghatNPB2026}. Nevertheless, it remains crucial to evaluate these models against current constraints on tidal deformability.

In the present work, we comprehensively explore the impact of DM on SQM in the CFL phase by contrasting a one--fluid model with a two--fluid framework. The fundamental distinction lies in their coupling mechanism: the one--fluid approach models the system via a unified EOS by assuming a microscopic Yukawa interaction between SQM and DM, whereas the two--fluid formalism treats the two components through independent EOSs coupled exclusively via gravity. We systematically analyze how varying {the ratio of central DM pressure to the total central pressure ($f_r$)}  and DM particle mass ($m_D$) modifies the stellar structural properties {(note that $f_r$ should not be interpreted as the actual DM mass fraction of the star)}. Specifically, we evaluate the mass--radius ($M-R$) and dimensionless tidal deformability--mass ($\Lambda-M$) relations to investigate how these observables respond to the underlying nature of the SQM--DM interaction. While the current observational baseline from GW170817 limits the parameter $\Lambda$ to $70 \lesssim \Lambda_{1.4 M_\odot} \lesssim 580$ ($90\%$ credible level) \cite{Abbott2018}, third--generation gravitational wave detectors---specifically the Einstein Telescope and Cosmic Explorer---are expected to suppress systematic errors and significantly refine these bounds \cite{Roy2024,Puecher2023}. Indeed, future empirical constraints are projected to tighten this threshold to $\Lambda_{1.4 M_\odot} \lesssim 400$, as suggested by \cite{Annala2018}. Motivated by these anticipated advancements in stellar radius and tidal deformability measurements, we systematically evaluate which coupling paradigm demonstrates better alignment with these projected constraints.

The structure of this paper is as follows. 
In Sec.~\ref{sec:EOS_one and two fluid}, we outline the formalism used to derive the thermodynamic potential and related properties--such as pressure, quark number densities, the EOS, and the sound speed--for SQSs admixed with scalar DM in both one--fluid and two--fluid models. Moreover, we discuss the constraints imposed on $m_D$ within the two--fluid framework.
Sec.~\ref{MR relation} presents the mass--radius ($M-R$) relations for various choices of $f_r$ and $m_D$, illustrating the behaviors in both the one--fluid and two--fluid models. In Sec.~\ref{tidal}, we outline the computation of the Love number alongside the dimensionless tidal deformability $\Lambda$, and examine the resulting $\Lambda$-M relations for different values of $f_r$ and $m_D$ within both frameworks. Finally, Sec.~\ref{sec:conclusion} provides a summary of our discussion and conclusions.

\section{Thermodynamic Properties of SQM Admixed with Scalar Dark Matter within One--and Two--Fluid Models} \label{sec:EOS_one and two fluid}

In this section, we outline the derivation of the thermodynamic potential of SQM, from which quark number densities, pressure and energy density are obtained. These quantities are then used to analyze the EOS, sound speed, and the causality condition of SQM in the presence of DM.
We begin with the one--fluid model, where SQM and DM are treated as a single interacting fluid, and subsequently, we turn to the two--fluid model, in which SQM and DM interact only gravitationally.
\subsection{Thermodynamic Properties of SQM in One--fluid Model}\label{thermodynamic properties}
In the one--fluid framework,  SQM and  DM are assumed to form a single, thermodynamically coupled fluid in full equilibrium.  
In the following, we outline the derivation of the thermodynamic potential for SQM coupled to scalar DM in the one--fluid model.

The total thermodynamic potential of the system is written as
\begin{equation}
	\Omega = \Omega_{\mathrm{free}} + \Omega_{\mathrm{QCD}} + \Omega_{\mathrm{CFL}} + \Omega_{\mathrm{Yukawa}} + \Omega_{\mathrm{BEC}}.
	\label{eq:Omega_total}
\end{equation}
In the above equation;
\begin{itemize}
	\item $\Omega_{\mathrm{free}}$ is the contribution from a free, relativistic Fermi gas of quarks and electrons \cite{Fraga2006,Kurkela2010}.
		\begin{eqnarray}
		-\frac{\Omega_{\mathrm{free}}}{V}
		= \sum_{N_f=1}^{3} \left( 	\frac{N_c \mu_f^4}{24\pi^2}\left( 2\hat{u_f}^3 - 3 z_f \hat{m_f}^2 \right)\right)
		+ \frac{\mu_e^4}{12\pi^2},
	\end{eqnarray}
	where
	\[
	\hat{u_f} \equiv \frac{\sqrt{\mu_f^2 - m_f^2}}{\mu_f}, \quad
	\hat{m_f} \equiv \frac{m_f}{\mu_f}, \]\\
	\[z_f \equiv \hat{u_f} - \hat{m_f}^2 \ln\!\left(\frac{1+\hat{u_f}}{\hat{m_f}}\right).
	\]
	 Here,  $N_f$ is the number of quark flavors, $N_C=3$ is the number of colors, $m_f$ and $\mu_f$ are the mass and chemical potential of quark flavor $f$, respectively, and $\mu_e$ is the electron chemical potential.
	 	\item $\Omega_{\mathrm{QCD}}$ represents the perturbative corrections from gluon--mediated quark--quark interactions, computed up to two--loop order in pQCD (see Fig. \ref{2loopQCD}) \cite{Fraga2006,Kurkela2010},
	\begin{equation}
		-\frac{\Omega_{\mathrm{QCD}}}{V}
		= \sum_{N_f=1}^{3}  \mathcal{F} \frac{\alpha_s(Q)}{4\pi},
		\end{equation}
where
		\begin{equation}
			\mathcal{F} =
			\frac{d_A \mu_f^4}{4\pi^2} \left( -6 z_f \hat{m_f}^2 \ln\frac{Q}{m_f}
			+ 2 \hat{u_f}^4 - 4 z_f \hat{m_f}^2 - 3 z_f^2 \right). \label{eq:m2}
		\end{equation}
$d_A \equiv N_c^2 - 1$ and $Q$ is the renormalization scale.
The running QCD coupling constant $\alpha_s$ is taken as \cite{Navas2025}
\begin{equation}
	\alpha_s(Q) =
	\frac{4 \pi}{\beta_0 L} \left(1 - \frac{2\beta_1 \ln L}{\beta_0^2 L}\right),
\end{equation}
where
\[
\beta_0 = 11 - \frac{2}{3}N_f, \quad
\beta_1 = 51 - \frac{19}{3}N_f, \]\\
\[ L = 2 \ln\!\left(\frac{Q}{\Lambda_{\overline{MS}}}\right).
\]
Here, the $\overline{MS}$ renormalization point, $\Lambda_{\overline{MS}}$, is extracted from the Particle Data Group (PDG) values \cite{Navas2025} by matching $\alpha_s(m_\tau)=0.314 $, with $m_\tau = 1776.86 \, \text{ MeV}$.
The quark sector consists of massless $u$ and $d$ quarks and a strange quark with a scale--dependent running mass $m_s(Q)$. The running mass is determined by the renormalization group equation \cite{Kurkela2010}, 
\begin{equation}
	m_s(Q) = m_s(2\,\mathrm{GeV})\left[ \frac{\alpha_s(Q)}{\alpha_s(2\,\mathrm{GeV})} \right]^{\gamma_0/\beta_0},
\end{equation}
where  $\beta_0$ and $\gamma_0$ are the standard one--loop beta and anomalous dimension coefficients, and $m_s(2\,\mathrm{GeV})=93.5 \text{ MeV}$ is taken from the latest PDG value \cite{Navas2025}.
{For additional details on the behavior of the running coupling and running mass, see Refs. \cite{Kurkela2010,sedaghatonefluid}.}
	\begin{figure}
		\center{\includegraphics[width=2.5cm]
			{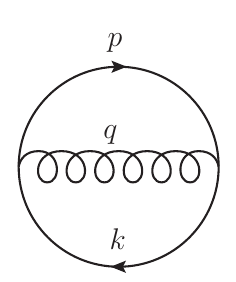}}
		\caption{\label{fig:one} \small{Two--loop diagram showing perturbative corrections arising from the gluon--mediated quark--quark interaction. Quark propagators are indicated by thick solid lines, and the gluon propagator is represented by a curly line.}}
		\label{2loopQCD}
	\end{figure}
\item $\Omega_{\mathrm{CFL}}$ represents the contribution from color superconductivity. 
At sufficiently high densities, quark matter is predicted to become a color superconductor, where quarks form Cooper pairs in analogy with electron pairing in conventional superconductors \cite{Alford2008, Rajagopal2001}. 
This pairing originates from the attractive nature of the strong interaction in specific color channels, leading to spontaneous color symmetry breaking and the onset of a superconducting phase \cite{Alford2008,Rajagopal2001}. 
Among the possible phases, the CFL state is favored at ultra--high densities, in which up, down, and strange quarks pair symmetrically, reducing the original SU(3) color and flavor symmetries to a common diagonal subgroup \cite{Shovkovy2005,Schmitt2006,Gholami2025}. 
The CFL phase has a profound impact on the EOS. The contribution of $\Omega_{\text{CFL}}$ is given by:
\begin{equation}
	-\frac{\Omega_{\text{CFL}}}{V}=\frac{\Delta^2 (\mu_u + \mu_d + \mu_s)^2}{3\pi^2},
\end{equation}
where $\Delta$ is the gap parameter. Astrophysical constraints on the dense--matter EOS, as noted in Ref.~\cite{Kurkela2024}, restrict the pairing gap to $\Delta \lesssim 216 \,\text{ MeV}$. In this work, we adopt $\Delta = 100 \,\text{ MeV}$.
Since quark pairing lowers the free energy while sustaining high pressure, the EOS becomes stiffer than that of unpaired quark matter. 
Such stiffening allows compact stars to sustain larger gravitational masses, with maximum values reaching $\sim 2 M_\odot$ in some models \cite{Kurkela2010,Sedaghat2021}. 
Although strongly supported by QCD theory, direct detection of color superconductivity is not feasible due to the extreme densities involved. 
Instead, indirect astrophysical signatures--such as neutron star cooling, rotational behavior, and neutrino emission patterns--may offer clues to the existence of this exotic phase \cite{Shovkovy2005,Schmitt2006,Page2009}.

\item $\Omega_{\mathrm{Yukawa}}$ denotes the contribution from the direct Yukawa interaction between quarks and scalar DM particles. 
This interaction is governed by the Lagrangian density
\begin{equation}
	\mathcal{L}_{\mathrm{int}} = - g \, \phi \, \bar{\psi} \psi, 
	\qquad g = \sqrt{4\pi\alpha_Y},
\end{equation}
where $\phi$ is the scalar DM field, $\psi$ the quark field, and $\alpha_Y$ the Yukawa coupling constant. 
The term $\Omega_{\mathrm{Yukawa}}$ is evaluated numerically through the two--loop diagram (Fig.~\ref{2loopYukawa}) 
in which a scalar propagator mediates the interaction between quark lines.  
\begin{figure}[htbp]
	\centering
	\includegraphics[width=2.5cm]{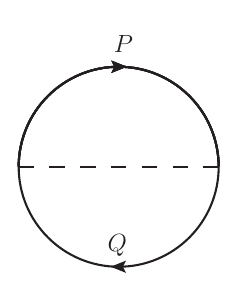}
	\caption{\small Two--loop diagram representing the Yukawa interaction between quarks and scalar DM. 
		The solid thick lines denote quark propagators, while the dashed line corresponds to the scalar DM propagator.}
	\label{2loopYukawa}
\end{figure}
The contribution of this diagram is computed using the finite--density cutting rules \cite{Ghisoiu2017}, 
which reduce the two--loop integrals to three--dimensional phase--space integrals over on--shell intermediate states. 
Further details of the calculation are provided in \cite{sedaghatonefluid}. 
This interaction modifies both the pressure and the energy density of the combined fluid, with its strength determined by $\alpha_Y$ and the DM particle mass $m_D$. 
In this work, we adopt $\alpha_Y = 0.05$.

\item $\Omega_{\mathrm{BEC}}$ denotes the mean--field contribution from a zero--temperature BEC phase of scalar DM, modeled as a dilute bosonic gas with short--range $s$--wave interactions. 
At zero temperature, scalar DM particles condense into a BEC once their thermal de Broglie wavelength surpasses the average interparticle distance. 
For a dilute, repulsive bosonic gas, $\Omega_{\mathrm{BEC}}$ can be expressed as
\begin{equation}\label{omegaBEC}
	-\frac{\Omega_{\mathrm{BEC}}}{V} = \frac{2 \pi l_a}{m_D^3} \epsilon_{\mathrm{BEC}}^2,
\end{equation}
{where $l_a$ denotes the $s$--wave scattering length which is typically set to $1\,\mathrm{fm}$ in compact star studies \cite{Panotopoulos2017,Lopes2018, X. Y. Li 2012} and $\epsilon_{\mathrm{BEC}}$ represents the condensate energy density (see \cite{Panotopoulos2017,Lopes2018,sedaghatonefluid} for further details). It is worth emphasizing that our chosen DM parameters are physically well--motivated. Specifically, in Ref.  \cite{Buras-Stubbs2024}, {a realistic parameter space for $m_D$ and $l_a$ has been established} by integrating observational constraints from galaxy cluster collisions  and the physics of compact stars, and our adopted values fall well within this viable regime. }
\end{itemize}

Having established the expression for  $\Omega$, we are now in a position to derive the corresponding macroscopic quantities of SQM. To ensure strict thermodynamic consistency \cite{Kurkela2010}, we first evaluate $n_B$ from the negative derivative of $\Omega$ with respect to the baryon chemical potential $\mu_B$:
\begin{equation}
	n_B = -\frac{1}{V} \left( \frac{\partial \Omega}{\partial \mu_B} \right),
\end{equation}
where the system is implicitly assumed to satisfy the requirements of the beta equilibrium and local charge neutrality \cite{Kurkela2010,Sedaghat2021,sedaghatannals,SedaghatNPB2026}.
The thermodynamic pressure $P$ is subsequently obtained by integrating this baryon density over the chemical potential:
\begin{equation}
	P = \int n_B \, d\mu_B - B_{\text{eff}},
\end{equation}
where the effective Bag constant, $B_{\text{eff}}$, naturally emerges as the integration constant. This parameter accounts for the non--perturbative QCD vacuum pressure and enforces the necessary boundary conditions for the quark phase. Without perturbative corrections, it reduces to the standard bag constant, while with corrections it acts as an effective parameter   \cite{Kurkela2010,Sedaghat2021,JSedaghat,sedaghat,sedaghatannals}.
Following this, the corresponding energy density $\epsilon$ is derived via the standard thermodynamic relation at zero temperature:
\begin{equation}
	\epsilon = -P + \mu_B n_B.
\end{equation}
By relating the pressure $P$ and the energy density $\epsilon$, the EOS for the DM--admixed SQM is self--consistently constructed. Finally, the speed of sound $c_s$ is evaluated from the resulting EOS as:
\begin{equation}
	c_s^2 = \frac{\partial P}{\partial \epsilon}.\label{cs}
\end{equation}
These thermodynamic inputs are essential for astrophysical applications, since the EOS directly enters the Tolman--Oppenheimer--Volkoff (TOV) equations, 
which determine the structural properties of compact stars such as their $M-R$ relation and stability. 
Thus, an accurate evaluation of $\Omega$ is a key step in connecting microscopic particle interactions with macroscopic stellar observables.
In this work, the effective bag constant is determined to be $B_{\text{eff}}=70 \text{ MeV/fm}^3$. 
The renormalization scale ($Q$) is chosen to be proportional to the average quark chemical potential \cite{Kurkela2010},
\begin{equation}
Q=X\,\bar{\mu},
\end{equation}
where $\bar{\mu}=(\mu_u+\mu_d+\mu_s)/3$ and $X$ is a dimensionless scale parameter \cite{Kurkela2010}. The parameter $Q$ is determined by imposing the Bodmer--Witten absolute stability condition, which requires the energy per baryon at zero pressure to be lower than $930\,\text{ MeV}$ \cite{bodmer,witten}. For $B_{\text{eff}} = 70 \text{ MeV/fm}^3$, the minimum value of the parameter $X$ satisfying absolute stability is $X \simeq 2.7$. Although larger 
values of $X$ remain stable, they suppress perturbative QCD 
corrections and diminish the sensitivity of the EOS to strong coupling 
effects. We therefore adopt $X = 2.7$ throughout this study. Apart from the Bodmer--Witten absolute stability condition, the EOS must satisfy several key viability criteria to ensure compliance with both fundamental physics and astrophysical observations. These requirements include:
\begin{itemize}
	\item Positive pressure throughout the stellar interior, with pressure vanishing at the surface.  
	\item Causality, guaranteed by the condition $c_s^2 = dP/d\epsilon \leq 1$.  
	\item Dynamical stability against radial oscillations, expressed through the adiabatic index $\Gamma = (dP/d\epsilon)(\epsilon + P)/P \geq 4/3$.  
\end{itemize}
In the following, we investigate how the EOS is controlled by different values of $f_r$ and $m_D$.  It is worth noting that, based on constraints on the DM self--interaction cross section derived in Refs.~\cite{Panotopoulos2017b,Lopes2018}, the DM particle mass is typically constrained to the range $50\,\mathrm{MeV} \leq m_D \leq 160\,\mathrm{MeV}$. However, other studies, such as Refs.~\cite{Karkevandi2022,SedaghatNPB2026,X.Li2012}, have explored larger values of $m_D$, extending up to $400\,\mathrm{MeV}$, $500\,\mathrm{MeV}$, and $1000\,\mathrm{MeV}$. In this work, we consider configurations that meet the $M_{\text{TOV}} > 2\,M_\odot$ threshold---for both the one--fluid and two--fluid models---to be consistent with the heavy pulsar PSR J0740+6620 ($M = 2.08 \pm 0.07 M_\odot$) \cite{Cromartie2019}. Since this pulsar provides one of the most reliable and precise lower bounds on the maximum mass of compact stars, this observational requirement leads us to adopt an upper bound of approximately $m_D \sim 400\,\mathrm{MeV}$. {Indeed, increasing $m_D$ beyond approximately $400\,\mathrm{MeV}$ does not allow most of the resulting configurations to satisfy the mass bound inferred from PSR J0740+6620. Therefore, we adopt $m_D \sim 400\,\mathrm{MeV}$ as the upper limit of the explored $m_D$ range.
	}
\begin{figure*}[htbp]
	\centering
	\par
	\includegraphics[width=17cm]{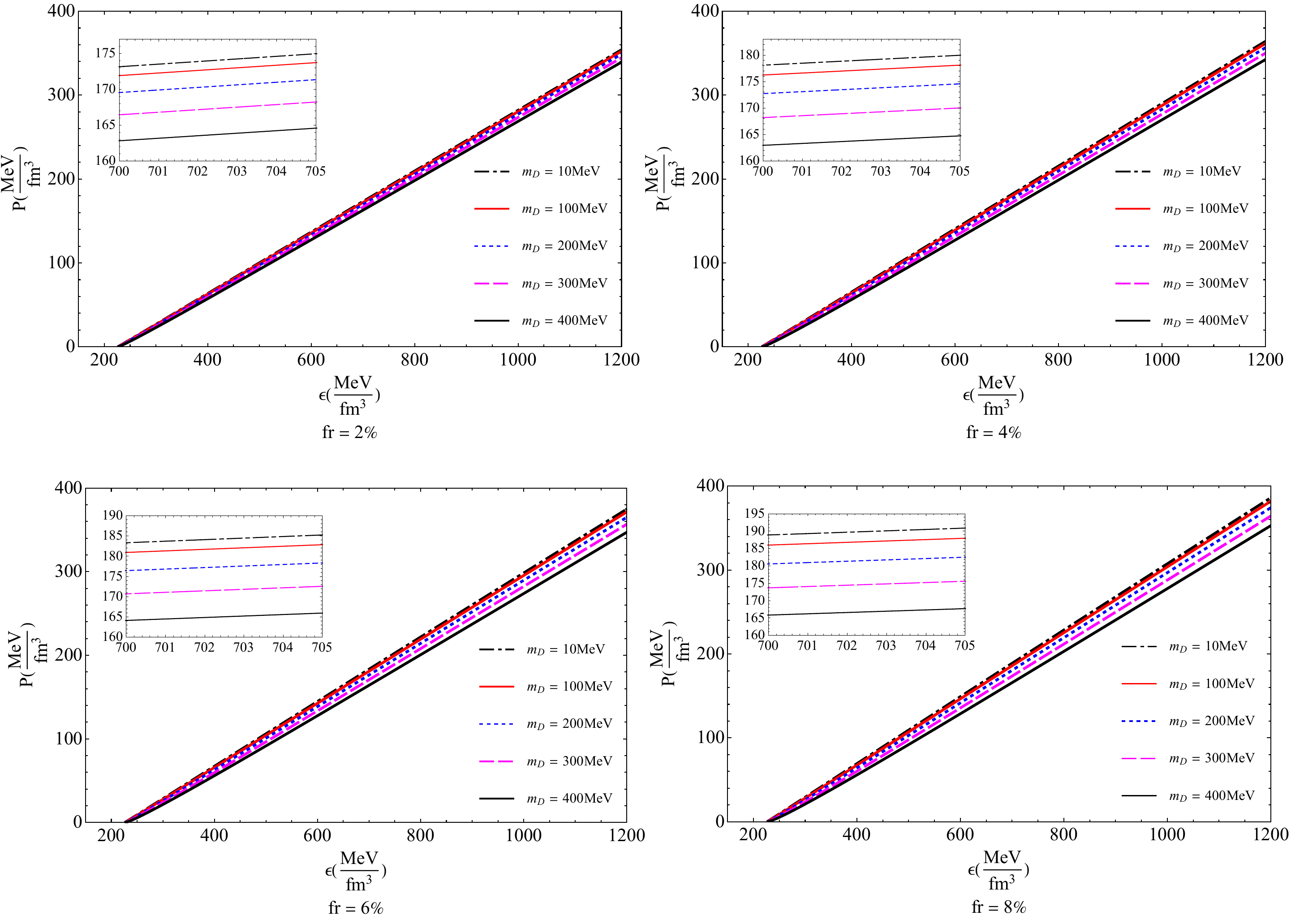}
	\caption{EOSs of SQM for different values of $f_r$ and $m_D$ in one--fluid model}
	\label{EOSs of one-fluid}
\end{figure*}

Figure~\ref{EOSs of one-fluid} presents EOS curves for different choices of $f_r$ and $m_D$ within the one--fluid framework. 
Two distinct trends emerge: at fixed $f_r$, increasing $m_D$ softens the EOS, as the pressure at a given energy density decreases, indicating reduced stiffness of the interacting mixture. 
In contrast, at fixed $m_D$, increasing $f_r$ stiffens the EOS, raising the pressure at a given energy density and reflecting stronger support from the larger fraction of interacting DM. 
These behaviors highlight how $m_D$ and $f_r$ jointly control the overall stiffness of the one--fluid EOS, driven by the competition between attractive Yukawa corrections and repulsive BEC pressure (for further discussion see Ref. \cite{sedaghatonefluid}). 
Such variations in the EOS directly impact other physical observable, notably the sound speed, the $M-R$ relation, and the tidal deformability parameter $\Lambda(M)$. 
The detailed consequences of these effects will be explored in the following.

\begin{figure*}[htbp]
	\centering
	\includegraphics[width=17cm]{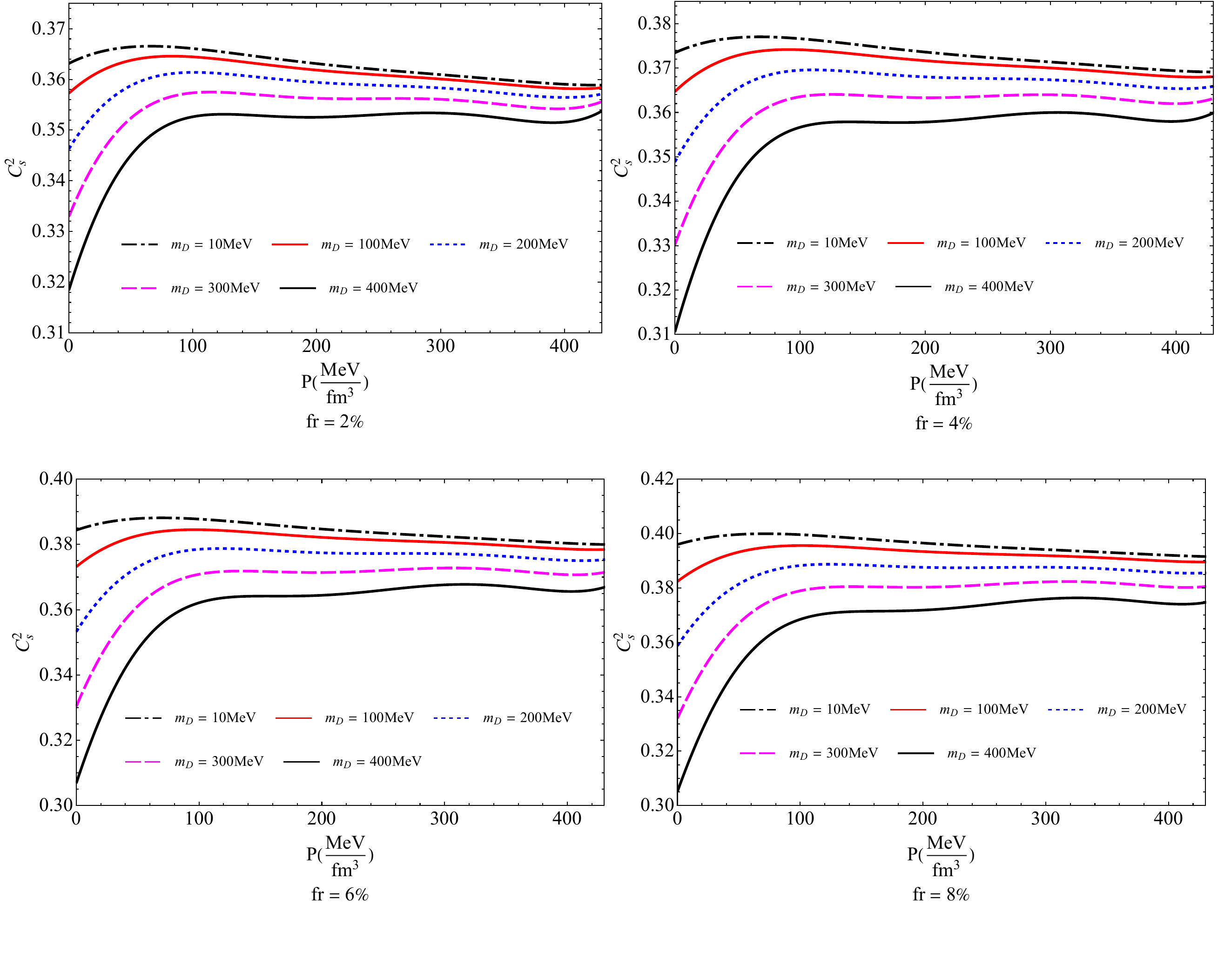}
	\caption{Sound speed squared $c_s^2$ in SQM as a function of pressure for different values of  $f_r$ and  $m_D$ in the one--fluid model. 
}
	\label{soundspeed}
\end{figure*}

The sound speed (Eq. (\ref{cs})) reflects the microscopic response of pressure to changes in energy density and provides a direct measure of the stiffness of the EOS. Since stiffness controls stellar stability and maximum mass, the sound speed serves as a useful diagnostic connecting microscopic interactions to macroscopic observables. 
As shown in Fig.~\ref{soundspeed}, the behavior of $c_s^2$ across different DM fractions $f_r$ and masses $m_D$ is fully consistent with the EOS trends discussed earlier. 
At fixed $f_r$, increasing the DM mass $m_D$ reduces the sound speed, reflecting the softening of the EOS due to the suppression of the BEC contribution. 
Conversely, at fixed $m_D$, increasing the fraction $f_r$ enhances the sound speed, indicating a stiffer EOS supported by a larger proportion of interacting DM. 
Importantly, in all cases the sound speed remains well below the causal limit $c_s^2 \leq 1$, confirming the physical consistency of the constructed EOSs. 

\begin{figure}[htbp]
	\centering
	\includegraphics[width=8cm]{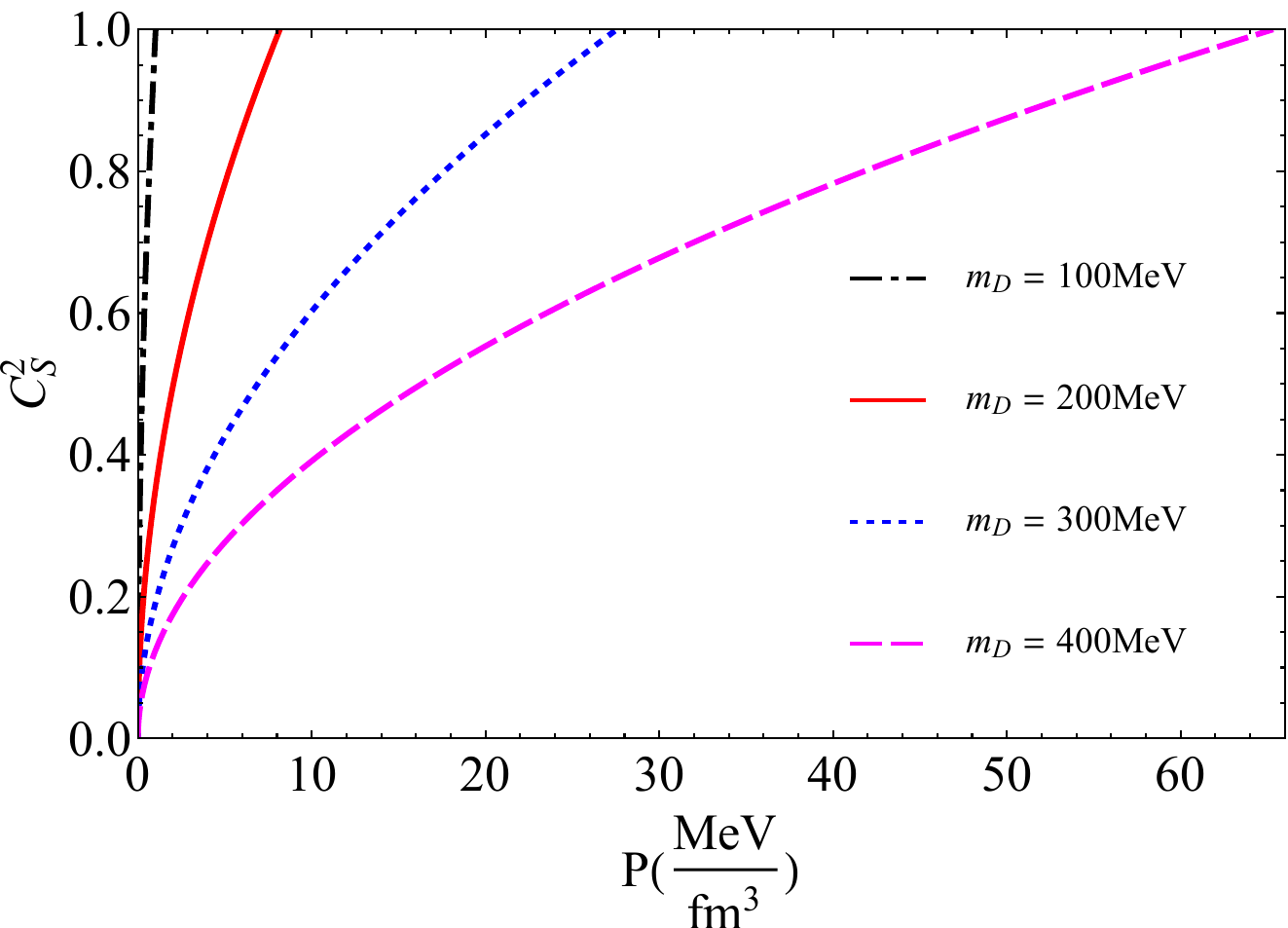}
	\caption{Sound speed squared $c_s^2$ in DM sector as a function of pressure for different values of $m_D$.}
	\label{fig:csDM}
\end{figure}
\begin{figure}[htbp]
	\centering
	\includegraphics[width=8cm]{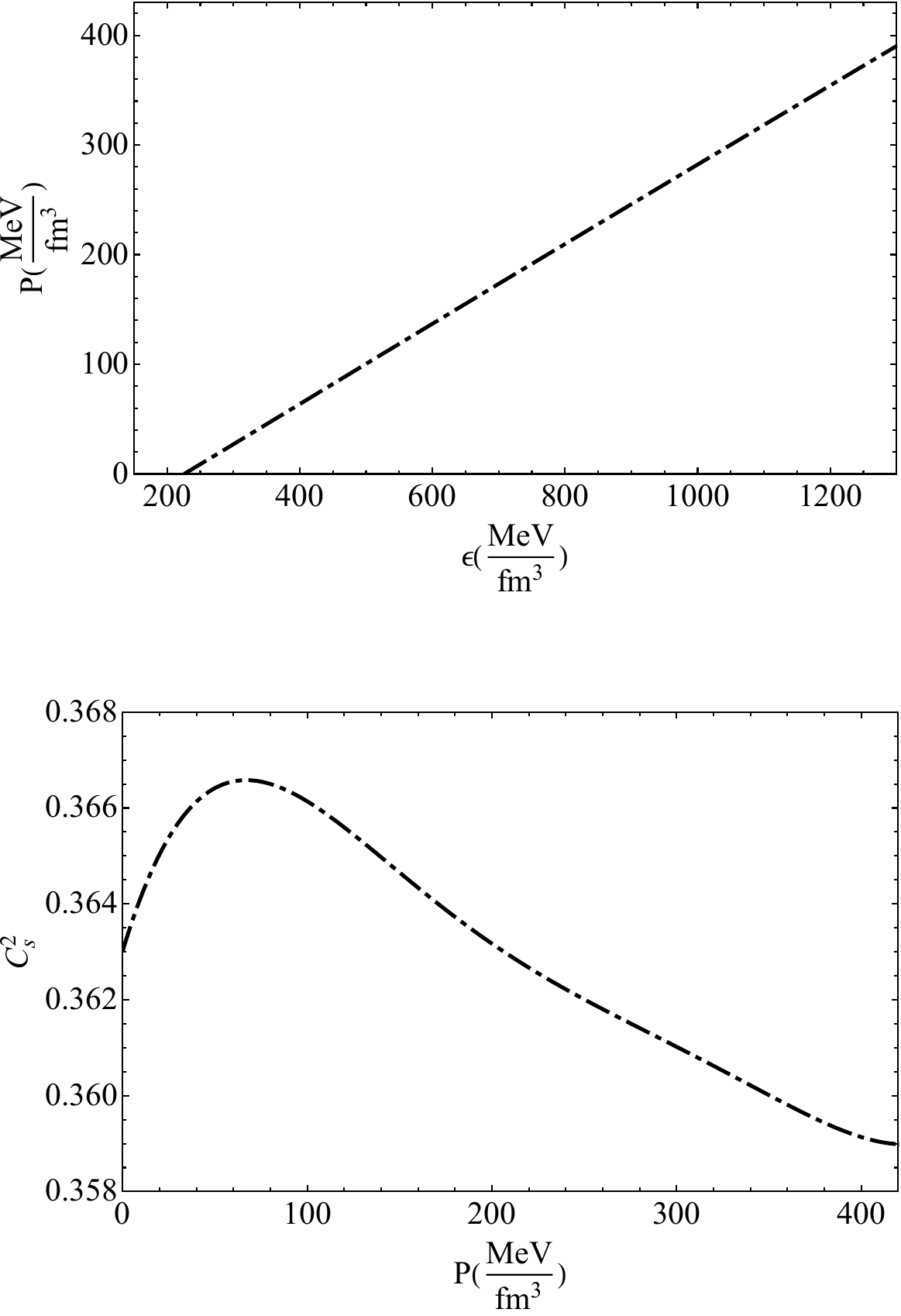}
	\caption{Equation of state (upper panel) and Sound speed squared $c_s^2$  (lower panel) for SQM in the two--fluid formalism.}
	\label{eossound2fluid}
\end{figure}

\subsection{Thermodynamic Properties of SQM in Two--Fluid Model} \label{Thermo-twofluid}

In the two--fluid framework, SQM and DM are treated as dynamically independent components that interact exclusively through gravity. In the absence of any non--gravitational coupling, each fluid obeys its own conservation laws and thermodynamic relations, which are determined solely by its underlying micro--physics and EOS. Since the sound speed is a local thermodynamic quantity defined by the response of pressure to changes in energy density, it depends only on the EOS of the corresponding fluid. Consequently, the DM component does not contribute to the local pressure response of SQM, and density perturbations in the quark matter sector do not induce a thermodynamic response in the DM sector. Therefore, the sound speed of SQM is entirely independent of the DM EOS, and vice versa.
As a result, the causality condition must be verified separately for each component.
This is in contrast to the one--fluid description, where SQM and DM are treated as a single effective medium. In the following, we first verify the causality condition in the DM sector.

From Eq.~(\ref{omegaBEC}), it is evident that the sound speed in DM sector increases linearly with the energy density. Such a behavior imposes a strong constraint on the maximum allowable energy density, and consequently on the maximum pressure, in the DM sector. This is because the linear growth of the sound speed tightens the causality bound $(c_s^2 \leq 1)$, restricting the admissible range of high--density configurations for the DM EOS. 
Figure~\ref{fig:csDM} displays $c_s^2$ as a function of pressure for different values of $m_D$. As can be seen, the value of $c_s^2$ increases monotonically with pressure and eventually reaches the causal limit, $c_s^2=1$. Consequently, the causality condition imposes an upper bound on the maximum pressure allowed in the DM sector $P^{DM}_{\rm max}$. Furthermore, increasing $m_D$ shifts the causal threshold to higher pressures, indicating that heavier DM particles remain causal over a wider density range. For instance, Fig.~\ref{fig:csDM} shows that $P^{DM}_{\rm max}\simeq 1.03~\mathrm{MeV/fm^3}$ for $m_D=100~\mathrm{MeV}$, whereas it increases to $P^{DM}_{\rm max}\simeq 27.44~\mathrm{MeV/fm^3}$ for $m_D=300~\mathrm{MeV}$. It should be emphasized that the parameter $f_r$ plays a crucial role in the present two--fluid model, as it dictates the central pressure fraction of the DM component. Consequently, the causality condition is inherently coupled to $f_r$ and must be verified for each specific value. This dependency significantly narrows the allowed range of $m_D$ in the two--fluid model compared to the one--fluid framework. Specifically, for any given $f_r$, the central DM pressure---which peaks at the stellar core---must not exceed the maximum threshold permitted by causality. Therefore, only configurations satisfying this condition are physically viable. As will be shown in the following section, the mass versus central DM pressure relation provides a convenient tool for identifying the maximum DM pressure reached inside the star and for determining the range of configurations that remain compatible with causality.

Accordingly, the thermodynamic potential of SQM is obtained from Eq.~(\ref{eq:Omega_total}) excluding the contributions of $\Omega_{\mathrm{Yukawa}}$ and $\Omega_{\mathrm{BEC}}$. The upper panel of Fig.~\ref{eossound2fluid} shows the pressure $P$ as a function of the energy density $\epsilon$, while the lower panel presents the corresponding squared sound speed as a function of pressure. Throughout the relevant density range, $c_s^2$ remains below the causal bound ($c_s^2 \leq 1$), confirming the physical consistency of the adopted EOS. 
\begin{figure*}[htbp]
	\centering
	\par
	\includegraphics[width=17cm]{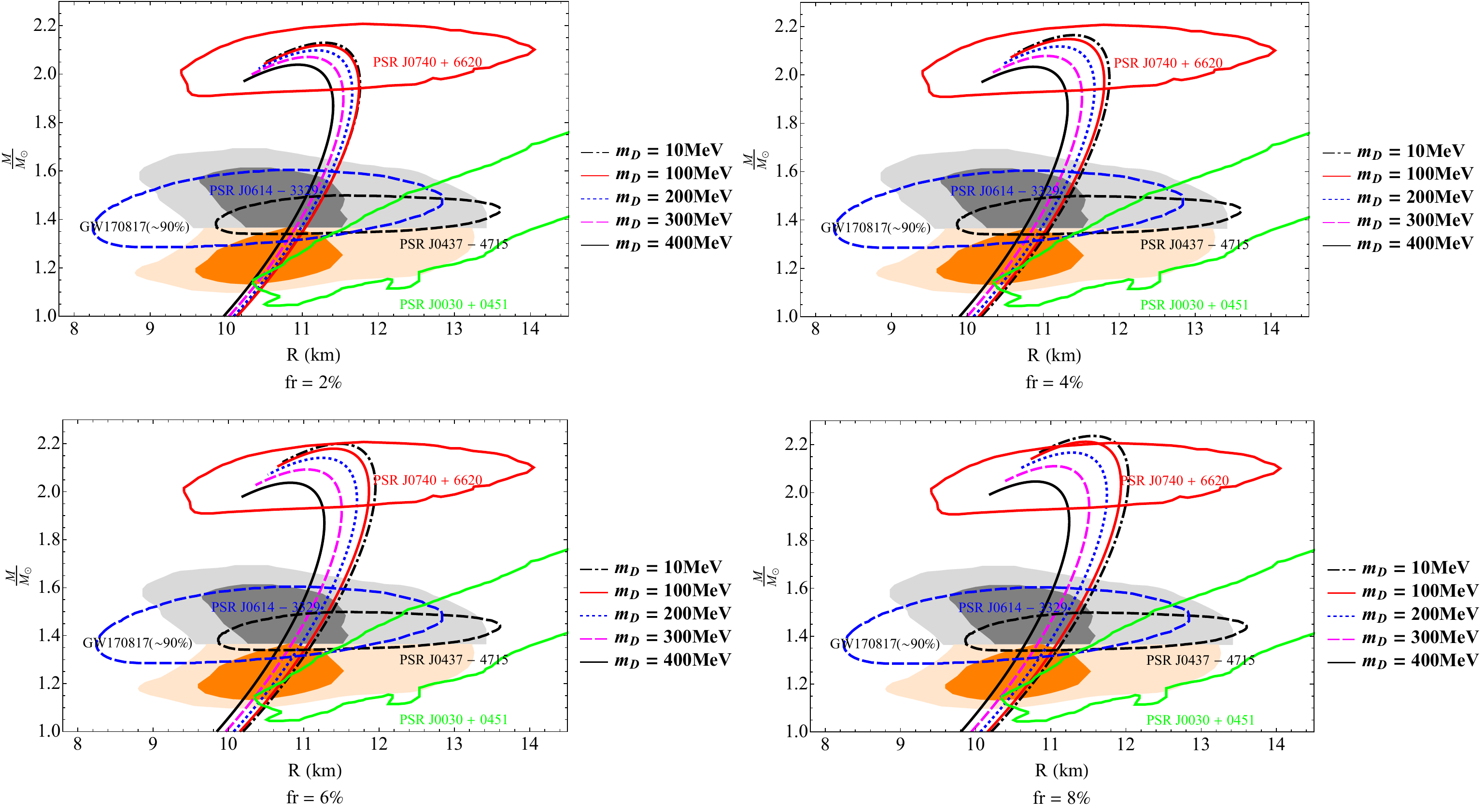}
	\caption{Mass--radius relations of one--fluid SQSs admixed with bosonic DM for different values of $m_D$ and $f_r$.  The green, blue, black, and red contours denote the observational mass--radius constraints for PSR~J0030$+$0451 \cite{Riley2019}, PSR~J0614$-$3329 \cite{PSR J0614 - 3329}, PSR~J0437$-$4715 \cite{PSR J0437 - 4715}, and PSR~J0740$+$6620 \cite{Cromartie2019}, respectively, while the gray and orange shaded regions correspond to the constraints inferred from the binary neutron-star merger GW170817 \cite{Abbott2017}.}
	\label{M-R-onefluid}
\end{figure*}

\section{Mass--radius relations for One-- and Two--Fluid Models} \label{MR relation}
The equilibrium structure of compact stars is determined by solving the Tolman--Oppenheimer--Volkoff (TOV) equations, which express the balance between pressure gradients and gravity in a spherically symmetric, static configuration within general relativity. The specific form of these equations depends on whether the stellar matter is modelled as a single interacting fluid or as two non--interacting fluids coupled only through gravity. In the following, we will present TOV equations for both approaches and illustrate the resulting mass--radius relations separately. 
\begin{table}[htbp]
	\centering
	\small
	\renewcommand{\arraystretch}{1.1} 
	\setlength{\tabcolsep}{5pt} 
	\caption{Structural properties of SQSs in one--fluid model for different values of $f_r$ and $m_D$.}
	\begin{tabular}{|c|c|c|c|c|}
		\hline
		\textbf{Model} & $f_r$ (\%) & $m_D$ (MeV) & $M_{\text{TOV}}$ ($M_\odot$) & $R$ (km) \\
		\hline
		&  & 10  & 2.13 & 11.29 \\
		&  & 100 & 2.12 & 11.27 \\
		One--Fluid & 2 & 200 & 2.10 & 11.19 \\
		&  & 300  & 2.07 & 11.07 \\
		&  & 400 & 2.04 & 10.95 \\
		\hline
		&  & 10  & 2.16 & 11.40 \\
		&  & 100 & 2.15 & 11.33 \\
		One--Fluid & 4 & 200 & 2.12 & 11.20 \\
		&  & 300  & 2.08 & 11.05 \\
		&  & 400 & 2.03 & 10.86 \\
		\hline
		&  & 10  & 2.20 & 11.48 \\
		&  & 100 & 2.18 & 11.40 \\
		One--Fluid & 6 & 200 & 2.14 & 11.24 \\
		&  & 300  & 2.09 & 11.04 \\
		&  & 400 & 2.04 & 10.82 \\
		\hline
		&  & 10  & 2.24 & 11.56 \\
		&  & 100 & 2.21 & 11.46 \\
		One--Fluid & 8 & 200 & 2.17 & 11.28 \\
		&  & 300  & 2.11 & 11.04 \\
		&  & 400 & 2.05 & 10.80 \\
		\hline
		
	\end{tabular}
	\label{tab:comparison_all}
\end{table}

\subsection{One--Fluid Mass--Radius Plots}

As previously mentioned in subsection~\ref{thermodynamic properties}, in the one--fluid framework, SQM and scalar DM are treated as a single thermodynamic system described by a unified EOS. The TOV equations take their standard form \cite{Tolman1939,Oppenheimer1939}:
\begin{align}
	\frac{dP(r)}{dr} &= - \frac{\left[ \epsilon(r) + P(r) \right] \left[ M(r) + 4\pi r^3 P(r) \right]}{r\left[ r - 2M(r) \right]}, \label{eq:TOV1} \\
	\frac{dM(r)}{dr} &= 4\pi r^2 \epsilon(r), \label{eq:TOV2}
\end{align}
where $M(r)$ is the enclosed gravitational mass within radius $r$, $P(r)$ and $\epsilon(r)$ are the total pressure and total energy density from the unified EOS (we adopt geometrized units in which $c=G=1$).
The integration starts from a central pressure $P_c$ (or equivalently a central energy density $\epsilon_c$) and $M(0) = 0$. The surface of the star, $R$, is reached when $P(R) = 0$, at which point $M(R)$ is the total gravitational mass $M_{\mathrm{TOV}}$. Stability is assessed via the condition $dM(R)/d\epsilon_c > 0$ \cite{Tangphati2021a,Tangphati2021b}.
Because the EOS already incorporates SQM and DM contributions, there is no need to evolve separate component profiles. All microscopic interactions, including the Yukawa coupling and BEC effects, are embedded in $P(\epsilon)$.
Figure~\ref{M-R-onefluid} and Table \ref{tab:comparison_all} show the $M-R$ relations for different choices of $f_r$ and $m_D$ in the one--fluid model. 
At fixed $f_r$, an increase in $m_D$ leads to a reduction in both $M_{\text{TOV}}$ and the corresponding radius.
{For example, at $f_r = 4\%$, increasing $m_D$ from $10$ to $400~\mathrm{MeV}$ reduces $M_{\text{TOV}}$ approximately from $2.16$ to $2.03\,M_\odot$, with the corresponding radius decreasing from $11.40$ to $10.86~\mathrm{km}$. This trend can be attributed to the fact that heavier DM softens the effective EOS when coupled to quark matter, reducing the pressure support against gravitational collapse.}
\begin{figure*}[htbp]
	\centering
	\par
	\includegraphics[width=18cm]{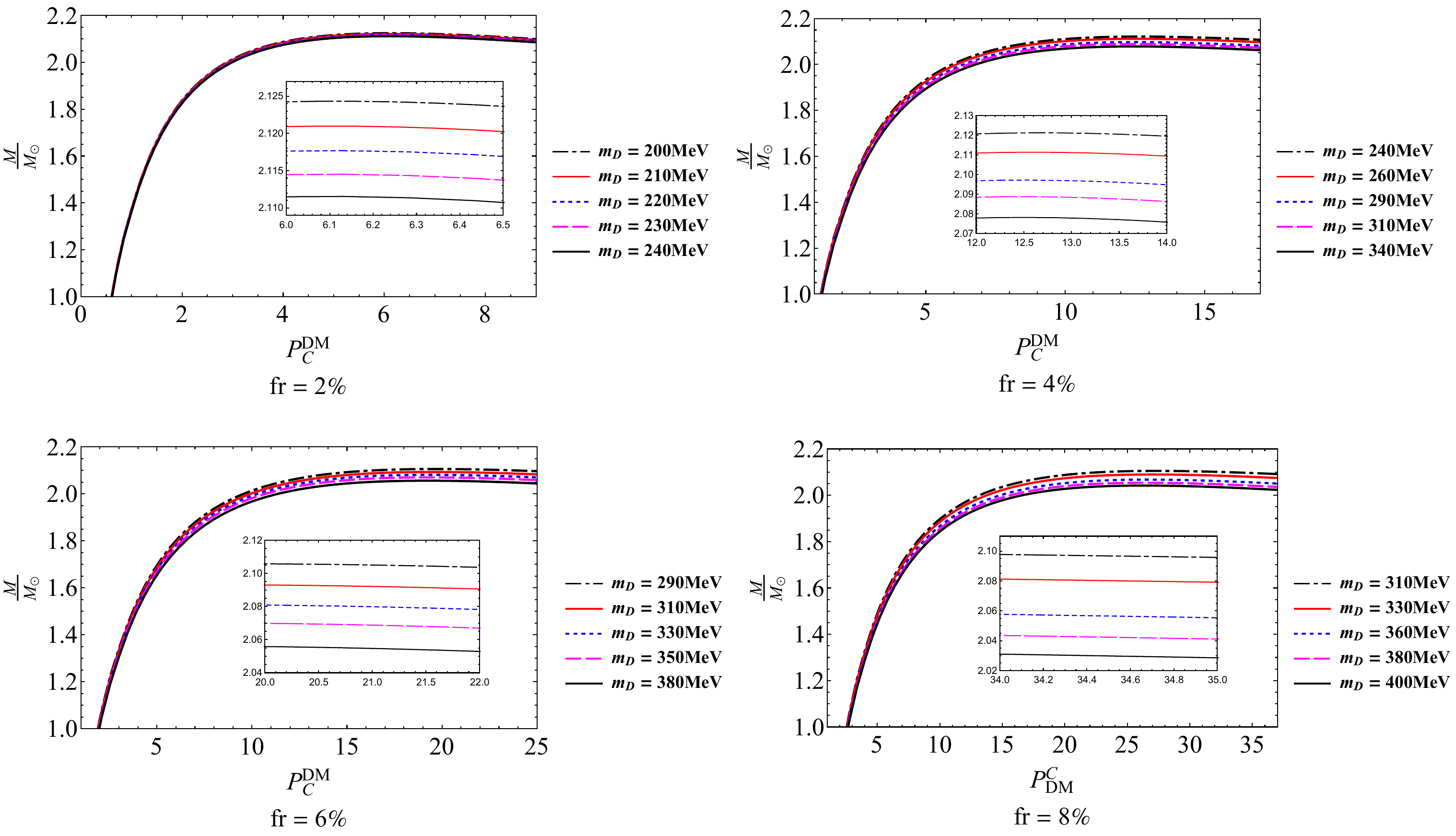}
	\caption{$M-P^{DM}_{C}$ diagrams for different values of $f_r$ and $m_D$ in two--fluid model.}
	\label{M-Pc-twofluid}
\end{figure*}
Conversely, at fixed $m_D$, increasing $f_r$ shifts the $M-R$ curve toward slightly larger radii and increases the maximum mass. {For instance, at $m_D=200~\mathrm{MeV}$, $M_{\text{TOV}}$ approximately rises  from $2.10\,M_\odot$ at $f_r=2\%$ to $2.17\,M_\odot$ at $f_r=8\%$, with a radius increase from $11.19$ to $11.28$~km.} This  behavior occurs because the interacting DM component contributes positively to the total pressure, effectively stiffening the EOS (see Fig. \ref{soundspeed}). Furthermore, the comparison with the observational $M-R$ constraints of pulsars shows that most of the considered configurations are compatible with the available measurements. However, a noticeable deviation appears for the case of $m_D=10~\mathrm{MeV}$. In this case, configurations with $f_r>2\%$ do not satisfy the constraints inferred from PSR~J0030$+$0451. This indicates that although increasing  $f_r$  can enhance the maximum mass of the star, it may simultaneously modify the radius in a way that becomes inconsistent with the corresponding observational limits. 
\subsection{Two--Fluid Mass--Radius Plots}
In the two--fluid picture, SQM and DM are treated as distinct fluids with separate EOSs, $P_{\mathrm{Q}}(\epsilon_{\mathrm{Q}})$ for quark matter and $P_{\mathrm{DM}}(\epsilon_{\mathrm{DM}})$ for DM. The two components interact only through the spacetime metric, i.e., gravitationally. Energy--momentum conservation holds separately for each fluid:
\begin{align}
	\nabla_\mu T^{\mu\nu}_{\mathrm{Q}} &= 0, \quad
	\nabla_\mu T^{\mu\nu}_{\mathrm{DM}} = 0.
\end{align}
This leads to a coupled set of TOV--like equations \cite{Panotopoulos2017,Karkevandi2022}:
\begin{align}
	&\frac{dP_{\mathrm{Q}}(r)}{dr}
	= -\big[\epsilon_{\mathrm{Q}}(r)+P_{\mathrm{Q}}(r)\big]
	\frac{M(r)+4\pi r^{3}P_{\mathrm{tot}}(r)}
	{r\big[r-2GM(r)\big]}, \\[1.5ex]
	&\frac{dP_{\mathrm{DM}}(r)}{dr}
	= -\big[\epsilon_{\mathrm{DM}}(r)+P_{\mathrm{DM}}(r)\big]
	\frac{M(r)+4\pi r^{3}P_{\mathrm{tot}}(r)}
	{r\big[r-2GM(r)\big]}, \\[1.5ex]
	&\frac{dM_{\mathrm{Q}}(r)}{dr}
	= 4\pi r^{2}\,\epsilon_{\mathrm{Q}}(r), \\[1.5ex]
	&\frac{dM_{\mathrm{DM}}(r)}{dr}
	= 4\pi r^{2}\,\epsilon_{\mathrm{DM}}(r), \\[1.5ex]
	&M(r)
	= M_{\mathrm{Q}}(r)+M_{\mathrm{DM}}(r), \\[1.5ex]
	&P_{\mathrm{tot}}(r)
	= P_{\mathrm{Q}}(r)+P_{\mathrm{DM}}(r), \\[1.5ex]
	&\epsilon_{\mathrm{tot}}(r)
	= \epsilon_{\mathrm{Q}}(r)+\epsilon_{\mathrm{DM}}(r).
\end{align}
The integration begins at the stellar center with given central pressures of SQM and DM. The radial profiles of the two components evolve until one of the pressures vanishes; the corresponding radius defines the surface of that fluid. In general, DM and SQM components do not necessarily share the same radius, potentially leading to a core--halo structure. {For comparison with the $M-R$ constraints obtained from NICER observations, we use the SQM (baryonic) radius, $R_{\mathrm{SQM}}$, which corresponds to the physical surface probed by the thermal X--ray emission from the pulsars. Since NICER measurements probe the emitting surface of the baryonic component, an extended DM halo can not be directly probed by these observations.}
\begin{figure*}
	\centering
	\par
	\includegraphics[width=18cm]{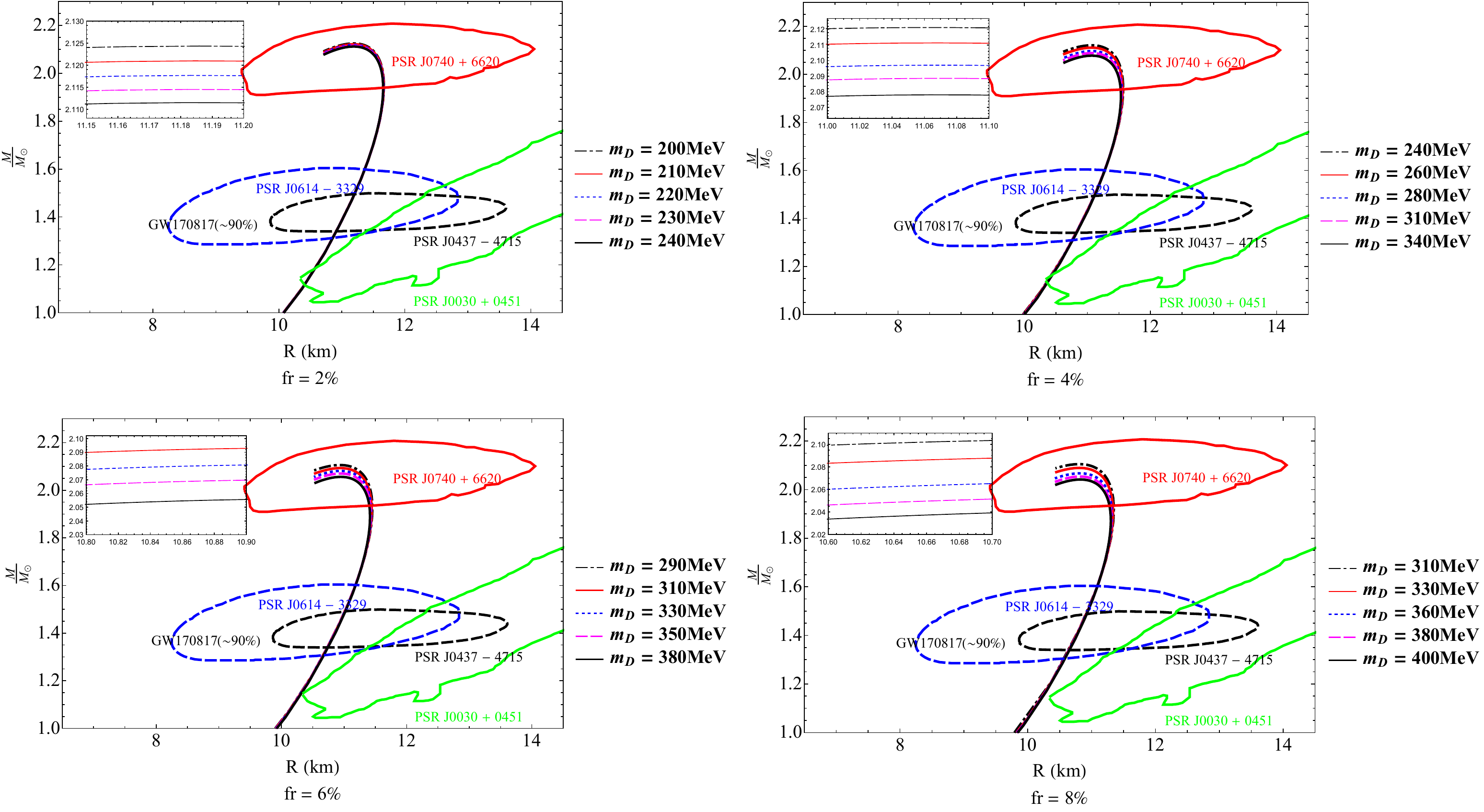}
	\caption{Mass--radius relations of two--fluid SQSs admixed with bosonic DM for different values of  $m_D$ and  $f_r$.  The green, blue, black, and red contours denote the observational mass--radius constraints for PSR~J0030$+$0451 \cite{Riley2019}, PSR~J0614$-$3329 \cite{PSR J0614 - 3329}, PSR~J0437$-$4715 \cite{PSR J0437 - 4715}, and PSR~J0740$+$6620 \cite{Cromartie2019}, respectively. Here, $R$ denotes the radius of the SQM component.}
	\label{M-R-twofluid}
\end{figure*}

As discussed in Subsection~\ref{Thermo-twofluid}, for each value of $m_D$ there exists a maximum DM pressure, $P^{DM}_{\rm max}$, beyond which the causality condition is violated. We also emphasized that the parameter $f_r$ plays a crucial role in the two--fluid model, as it determines the central DM pressure, $P^{DM}_{C}$, within the star. Consequently, physically acceptable stellar configurations must satisfy the condition $P^{DM}_{C}<P^{DM}_{\rm max}$. Therefore, the causality constraint must be examined separately for each adopted combination of $f_r$ and $m_D$.
To this end, for every value of $f_r$, we considered different values of $m_D$ and investigated the $M-P^{DM}_{C}$ relation in order to determine the central DM pressure corresponding to  $M_{\rm TOV}$. Since this configuration represents the highest $P^{DM}_{C}$ reached along a given stellar sequence, we denote it by $P^{DM}_{C,\rm max}$. The obtained values of $P^{DM}_{C,\rm max}$ were then compared with the corresponding causal limits, $P^{DM}_{\rm max}$. Parameter sets satisfying the condition $P^{DM}_{C,\rm max}<P^{DM}_{\rm max}$ were regarded as physically admissible, whereas those violating this requirement were excluded. In this way, a lower bound on $m_D$ is established for each adopted value of $f_r$. To illustrate this procedure more clearly, we present the $M-P^{DM}_{C}$ relations in Fig.~\ref{M-Pc-twofluid} for different values of $m_D$ and $f_r$. {Note that while $M$ is conventionally plotted against the total central energy density, here we deliberately present $M$ as a function of $P^{DM}_{C}$ to  explicitly highlight the threshold behavior and isolate $P^{DM}_{\rm max}$ prior to the onset of TOV instability.} It can be seen that, for each value of $f_r$, a minimum allowed value of $m_D$ can be identified by requiring that the condition $P^{DM}_{C,\rm max}<P^{DM}_{\rm max}$ remains satisfied.
As an example, for $f_r=2\%$, the minimum allowed $m_D$ is found to be approximately $200~\mathrm{MeV}$. According to Fig.~\ref{fig:csDM}, the corresponding causal upper limit is $P^{DM}_{\rm max}\simeq 8~\mathrm{MeV/fm^3}$, whereas Fig.~\ref{M-Pc-twofluid} indicates that  $P^{DM}_{C,\rm max}\simeq 6.5~\mathrm{MeV/fm^3}$. Since  the condition $P^{DM}_{C,\rm max}<P^{DM}_{\rm max}$ is fulfilled, the corresponding stellar configuration is considered physically admissible. Therefore, in the following, we investigate the structural properties of two--fluid SQSs using the values of $f_r$ and $m_D$ shown in Fig.~\ref{M-Pc-twofluid}.

{Figure~\ref{M-R-twofluid} and Table~\ref{tab:comparison_all2} present the $M-R$ relations obtained within the two--fluid framework. It should first be noted that the radius plotted in these configurations corresponds  to the baryonic (SQM) radius, which is the physical boundary probed by pulsar thermal emission observations (such as NICER). In contrast, the observational constraints derived from the gravitational waves of GW170817 are inherently sensitive to the total radius of the dark matter--admixed system. In the two--fluid configurations, this total radius is determined by the outermost boundary of the system, and is therefore given by
	$R=\max(R_{\rm SQM},R_{\rm DM})$.
	Consequently, the SQM radius is used when comparing the model predictions with the NICER mass--radius constraints, whereas the total radius is used in the calculation of the compactness, Love number, and tidal deformability relevant to the gravitational wave response.
Therefore, the constraints from GW170817 are omitted from these specific diagrams to ensure a physically consistent comparison.} Comparison with the observational $M-R$ constraints indicates that almost all considered configurations remain compatible with the available pulsar measurements. The only exception is the case with $f_r=8\%$, for which the predicted $M-R$ relation falls outside the confidence region inferred for PSR~J0030$+$0451, whereas it remains consistent with the constraints from the other observed pulsars. {Since our aim is to investigate configurations that remain as consistent as possible with the available observational $M-R$ constraints, we restrict the explored range to $f_r\leq8\%$. We include $f_r=8\%$ to investigate the effects of a relatively large DM contribution and  facilitate a consistent comparison between the one--fluid and two--fluid models.}
The results also indicate that the two--fluid model exhibits the same qualitative trends as the one--fluid description. For a fixed value of $f_r$, increasing $m_D$ reduces both the $M_{\rm TOV}$ and the corresponding stellar radius. Conversely, for a fixed value of $m_D$, larger values of $f_r$ lead to higher values of $M_{\rm TOV}$ and $R$.
Despite these similarities, the dependence of the stellar properties on the DM parameters is significantly weaker in the two--fluid framework. As shown in Figs.~\ref{M-R-onefluid} and \ref{M-R-twofluid}, as well as in Tables~\ref{tab:comparison_all} and \ref{tab:comparison_all2}, the variations of both $M_{\rm TOV}$ and $R$ occur more gradually in the two--fluid model. This behavior can be attributed to the different roles played by DM in the two approaches. In the one--fluid description, DM directly modifies the effective EOS of the stellar medium, making the maximum mass and radius highly sensitive to variations in the DM parameters. In contrast, within the two--fluid framework, SQM and DM are treated as independent components that interact only through gravity. Consequently, changes in the DM sector influence $M_{\rm TOV}$ and $R$ only indirectly through the gravitational field, leading to a much smoother evolution of the macroscopic stellar properties.
\begin{table}[htbp]
	\centering
	\small
	\renewcommand{\arraystretch}{1.1} 
	\setlength{\tabcolsep}{5pt} 
	\caption{Structural properties of SQSs in two--fluid model for different values of $f_r$ and $m_D$. Here, $R$ denotes the radius of the SQM component.}
	\begin{tabular}{|c|c|c|c|c|}
		\hline
		\textbf{Model} & $f_r$ (\%) & $m_D$ (MeV) & $M_{\text{TOV}}$ ($M_\odot$) & $R$ (km) \\
		\hline
		&  & 200  & 2.12 & 11.19 \\
		&  & 210 & 2.12 & 11.18 \\
		Two--Fluid & 2 & 220 & 2.12 & 11.18 \\
		&  & 230  & 2.11 & 11.18 \\
		&  & 240 & 2.11 & 11.18 \\
		\hline
		&  & 240 & 2.12 & 11.07 \\
		&  & 260 & 2.11 & 11.07 \\
		Two--Fluid & 4 & 280 & 2.10 & 11.06 \\
		&  & 310  & 2.09 & 11.06 \\
		&  & 340 & 2.08 & 11.06 \\
		\hline
		&  & 290  & 2.11 & 10.93 \\
		&  & 310 & 2.10 & 10.95 \\
		Two--Fluid & 6 & 330 & 2.08 & 10.94 \\
		&  & 350  & 2.07 & 10.94 \\
		&  & 380 & 2.06 & 10.94 \\
		\hline
		&  & 310 & 2.11 & 10.80 \\
		&  & 330 & 2.09 & 10.81 \\
		Two--Fluid & 8 & 360 & 2.07 & 10.83 \\
		&  & 380  & 2.05 & 10.82 \\
		&  & 400 & 2.04 & 10.82 \\
		\hline
	\end{tabular}
	\label{tab:comparison_all2}
\end{table}
\section{Dimensionless Tidal Deformability in One--Fluid and Two--Fluid Models} \label{tidal}

The tidal deformability of a compact star quantifies the degree to which the star's shape is distorted by the tidal field of a binary companion. It is an observable of high astrophysical significance, as it imprints directly on the gravitational wave signal during the late inspiral phase of binary mergers. Measurements from events such as GW170817 have placed stringent upper bounds on the tidal deformability of a $1.4\,M_\odot$ star, with $\Lambda_{1.4 M_\odot}\lesssim 580$ at 90\% confidence. This constraint is a powerful tool for testing EOSs and discriminating between proposed models of dense matter.
For a static, spherically symmetric star in hydrostatic equilibrium, the dimensionless tidal deformability $\Lambda$ is defined by
\begin{equation}
	\Lambda \equiv \frac{2}{3} k_2 C^{-5}, \quad C \equiv \frac{M}{R},
\end{equation}
where $k_2$ is the quadrupolar ($l=2$) tidal Love number, $M$ is the gravitational mass, $R$ is the stellar radius, and $C$ is the compactness parameter (in geometrized units).
The Love number $k_2$ encodes the linear response of the star's mass quadrupole to an external tidal field and is given by \cite{Hinderer2009,Hinderer2010,Postnikov2010}:
\begin{align}
	k_2 &= \frac{16}{15} C^5 (1 - 2C)^2 \left[ 2 + 2C(y_R - 1) - y_R \right] \nonumber \\
	&\quad \times \Bigg\{ 2C \left[ 6 - 3y_R + 3C(5y_R - 8) \right] \nonumber \\
	&\qquad + 4C^3 \left[ 13 - 11y_R + C(3y_R - 2) + 2C^2(1 + y_R) \right] \nonumber \\
	&\qquad + 3(1 - 2C)^2 \left[ 2 - y_R + 2C(y_R - 1) \right] \ln(1 - 2C) \Bigg\}^{-1}
	\label{eq:k2}
\end{align}
Here, the effective surface value $y_R$ is determined by adding a surface correction to the interior value:
\begin{equation}\label{yR for SQS}
y_R = y(R) - \frac{4\pi R^3 \epsilon_0}{M},
\end{equation}
where the second term is a correction specific to SQSs due to the sudden drop of energy density at the boundary ($\epsilon_0$ is the energy density just inside the surface) \cite{Hinderer2010}. On the other hand, the first term, $y(R)$, is obtained by integrating the following differential equation from the center to the surface \cite{Postnikov2010}:
\begin{align}\label{eq:y_evol}
r y'(r) &+ y(r)^2 + r^2 Q(r) \nonumber \\
&+ y(r) e^{\lambda(r)} \left[ 1 + 4\pi r^2 (p(r) - \epsilon(r)) \right] = 0.
\end{align}
The interior coefficient $Q(r)$ is defined as:
\begin{align}
Q(r) &= 4\pi e^{\lambda(r)} \left( 5\epsilon(r) + 9p(r) + \frac{\epsilon(r) + p(r)}{c_s^2(r)} \right) \nonumber \\
&\quad - 6 \frac{e^{\lambda(r)}}{r^2} - (\nu'(r))^2,
\end{align}
where the required metric functions are given by:
\begin{equation}
e^{\lambda(r)} = \left[ 1 - \frac{2M(r)}{r} \right]^{-1},
\end{equation}
\begin{equation}
\nu'(r) = 2e^{\lambda(r)} \frac{M(r) + 4\pi p(r) r^3}{r^2}.
\end{equation}
The numerical integration starts from the center ($r=0$) with the standard initial condition $y(0)=2$ and proceeds outward to the stellar surface, where the surface value $y_R=y(R)$ is obtained. 
\subsection{One--Fluid Model}
\begin{figure*}
	\centering
	\includegraphics[width=14.6cm]{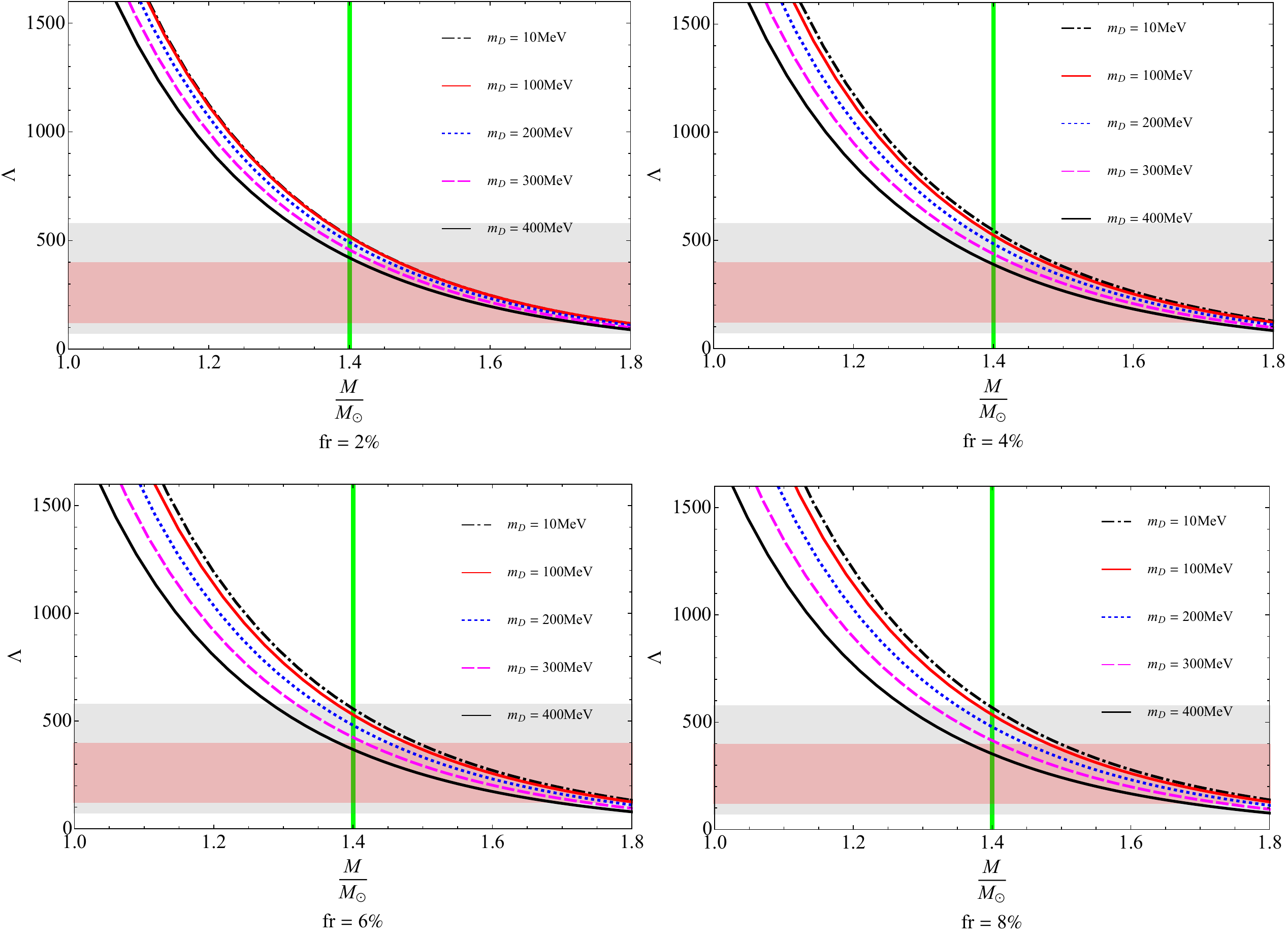}
	\caption{$\Lambda$--$M$ relations for different values of $f_r$ and $m_D$ in the one--fluid model. The gray shaded region represents the current limit
		from GW170817, $\Lambda_{1.4 M_\odot}\lesssim 580$, while the red shaded region
		indicates the anticipated future bound, $\Lambda_{1.4 M_\odot}\lesssim 400$.}
	\label{tidal-onefluid}
\end{figure*}

In the one--fluid formalism, the total pressure $P(r)$ and energy density $\epsilon(r)$ obtained from the unified EOS enter directly into the TOV equations (Eqs. (\ref{eq:TOV1}) and (\ref{eq:TOV2})) and the perturbation equation (Eq. (\ref{eq:y_evol})). 
The resulting values of $y_R$ and compactness $C$ are then substituted into Eq.~(\ref{eq:k2}) to compute  $k_2$ and  $\Lambda$. 
Figure~\ref{tidal-onefluid} displays $\Lambda$ as a function of $M/M_\odot$ for our one--fluid model. We investigate this relationship across different values of $m_D$ and $f_r$. To evaluate our results against observational limits, we consider two areas at the canonical mass of $1.4\,M_\odot$ (marked by the vertical green line). The first area is the gray band ($70 \le \Lambda_{1.4 M_\odot} \le 580$), which represents the current constraint from the GW170817 event by the LIGO/Virgo collaboration \cite{Abbott2018}. The second is the tighter pink band ($120 \le \Lambda_{1.4 M_\odot}\le 400$), which was proposed as a prospective constraint by combining observational data with theoretical modeling \cite{Annala2018}. Specifically, the upper limit of $\Lambda_{1.4 M_\odot} \le 400$ corresponds to the tighter $50\%$ credible interval of the GW170817 event \cite{Abbott2017}. Meanwhile, the lower bound of $\Lambda_{1.4 M_\odot} \ge 120$ is a theoretical limit obtained by imposing strict causality ($c_s^2 \le 1$) and requiring the maximum TOV mass to exceed $2.0\,M_\odot$ within state--of--the--art equation of state (EOS) interpolations \cite{Annala2018}. 
Based on the analysis in \cite{Annala2018}, tightening the $\Lambda$ limit to $\Lambda_{1.4 M_\odot} \le 400$ directly restricts the radius of a $1.4\,M_\odot$ star to $R_{1.4 M_\odot} < 12.5\text{ km}$. In fact, recent stringent multi--messenger analyses and nuclear theory constraints suggest an even smaller upper bound of $11.9\text{ km}$ \cite{Capano2020}. Therefore, this range for $\Lambda$ (and the resulting small stellar radii) is expected to be directly tested by future high--precision astronomical observations, such as advanced X--ray missions and next--generation gravitational wave detectors.
Our analysis reveals that while the broader constraint of $\Lambda_{1.4 M_\odot} \le 580$ is easily satisfied across  the parameter space, the prospective $\Lambda_{1.4 M_\odot} \le 400$ limit is met under much more restricted conditions. Satisfying this prospective future constraint requires a specific choice of parameters. Specifically, it requires a higher DM fraction ($f_r \ge 6\%-8\%$) alongside heavier dark particles ($m_D \ge 200\text{ MeV}$), as also compared in Table~\ref{tab:comparison_tidal}. In the next step of our study, we turn off the direct interaction between the DM and the SQM, thereby transitioning our framework from a one--fluid model to a two--fluid system. We then investigate how this structural modification alters the $\Lambda-M$ behavior.
\begin{table}
	\centering
	\small
	\renewcommand{\arraystretch}{0.97} 
	\setlength{\tabcolsep}{5pt} 
	\caption{Comparison of $\Lambda_{1.4 M_\odot}$ values in one--fluid model for different values of $f_r$ and $m_D$.}
	\begin{tabular}{|c|c|c|c|}
		\hline
		\textbf{Model} & $f_r$ (\%) & $m_D$ (MeV) & $\Lambda_{1.4 M_\odot}$ \\
		\hline
		&  & 10  & 518.24  \\
		
		&  & 100  & 517.80 \\
		
		One--Fluid & 2 & 200  & 503.97  \\
		
		&  & 300  & 457.32 \\
		
		&  & 400  & 420.55  \\
		\hline
		&  & 10  & 522.02  \\
		
		&  & 100  & 521.69  \\
		
		One--Fluid & 4 & 200  & 483.58  \\
		
		&  & 300  & 438.29  \\
		
		&  & 400  & 390.01  \\
		\hline
		&  & 10  & 557.42  \\
		
		&  & 100  &  530.07 \\
		
		One--Fluid & 6 & 200  & 481.81  \\
		
		&  & 300  & 423.85  \\
		
		&  & 400  & 369.01  \\
		
		\hline
		&  & 10  & 568.29  \\
		
		&  & 100  & 534.04  \\
		
		One--Fluid & 8 & 200  & 476.89  \\
		
		&  & 300  & 413.21  \\
		
		&  & 400  & 351.24  \\
		\hline
	\end{tabular}
	\label{tab:comparison_tidal}
\end{table}
\begin{figure*}
	\centering
	\includegraphics[width=14.5cm]{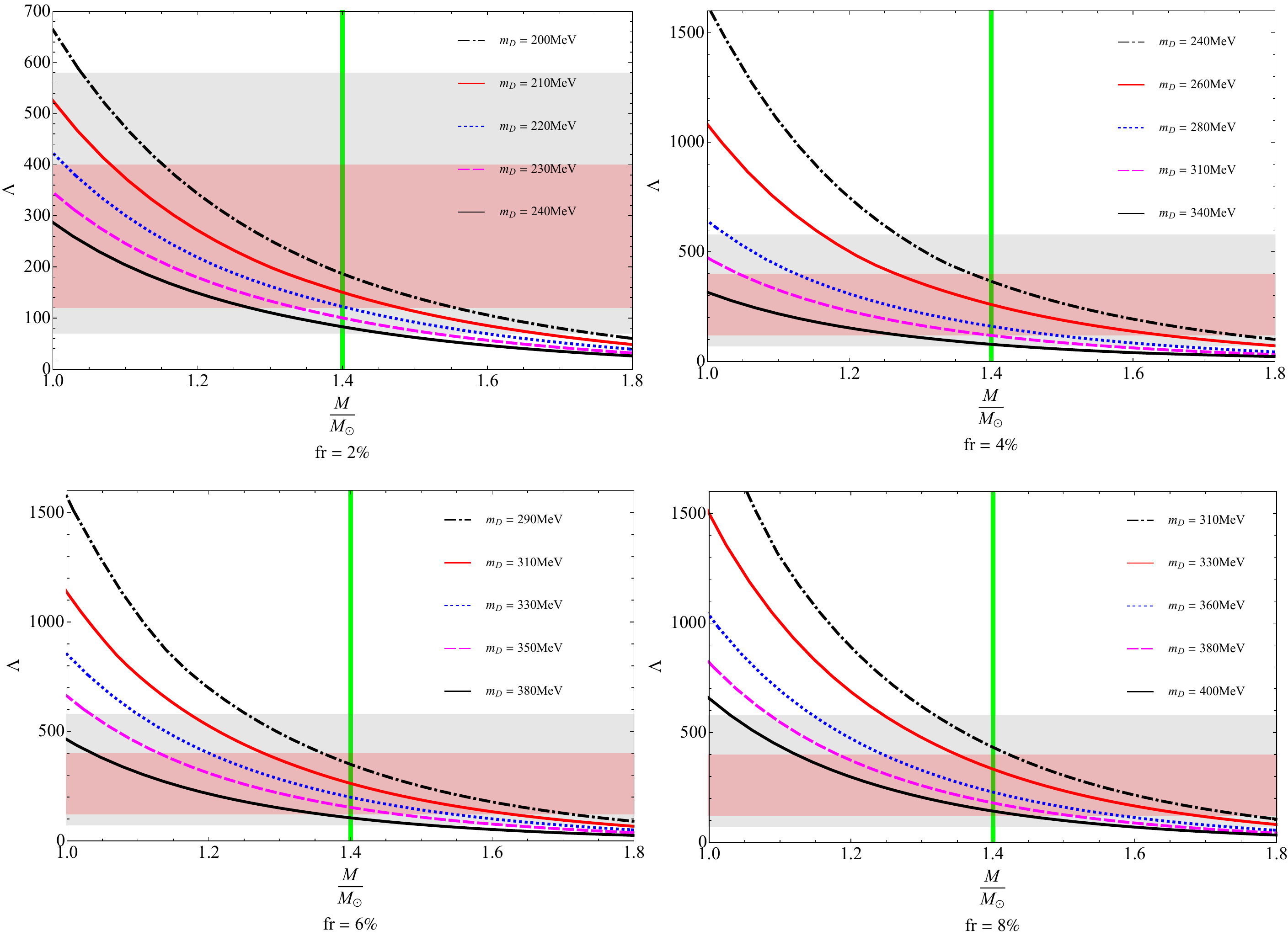}
	\caption{$\Lambda$--$M$ relations for different values of $f_r$ and $m_D$ in the two--fluid model. The gray shaded region represents the current limit
		from GW170817, $\Lambda_{1.4 M_\odot}\lesssim 580$, while the red shaded region
		indicates the anticipated future bound, $\Lambda_{1.4 M_\odot}\lesssim 400$.}
	\label{tidal-twofluid}
\end{figure*}
\begin{table}
	\centering
	\small
	\renewcommand{\arraystretch}{0.98} 
	\setlength{\tabcolsep}{5pt} 
	\caption{Comparison of $\Lambda_{1.4 M_\odot}$ values in two--fluid model for different values of $f_r$ and $m_D$.}
	\begin{tabular}{|c|c|c|c|}
		\hline
		\textbf{Model} & $f_r$ (\%) & $m_D$ (MeV) & $\Lambda_{1.4 M_\odot}$ \\
		\hline
		&  & 200  & 187.08  \\
		
		&  & 210  & 150.38  \\
		
		Two--Fluid & 2 & 220  & 122.73  \\
		
		&  & 230  & 101.36  \\
		
		&  & 240  & 82.83 \\
		\hline
		&  & 240  & 364.99  \\
		
		&  & 260  & 261.03  \\
		
		Two--Fluid & 4 & 280  & 160.81  \\
		
		&  & 310  & 119.30  \\
		
		&  & 290  & 78.43  \\
		\hline
		&  & 290  & 350.92  \\
		
		&  & 310  & 263.23  \\
		
		Two--Fluid & 6 & 330  & 199.81  \\
		
		&  & 350  & 153.50  \\
		
		&  & 380  & 104.97  \\
		
		\hline
		&  & 310  & 434.16  \\
		
		&  & 330  & 333.30  \\
		
		Two--Fluid & 8 & 360  & 228.73  \\
		
		&  & 380  & 179.61  \\
		
		&  & 400  & 142.33  \\
		\hline
	\end{tabular}
	\label{tab:comparison_tidal2}
\end{table}
\subsection{Two-Fluid Model}

In the two--fluid framework, Eq. \eqref{eq:y_evol} is modified to consider the total pressure and energy density as the sums over the SQM and DM components:
\begin{align}
	P_{\mathrm{tot}}(r) &= P_{\mathrm{Q}}(r) + P_{\mathrm{DM}}(r), \\
	\epsilon_{\mathrm{tot}}(r) &= \epsilon_{\mathrm{Q}}(r) + \epsilon_{\mathrm{DM}}(r).
\end{align}
The coupled sets of TOV equations (Sec.~\ref{MR relation}) are solved simultaneously along with a single $y(r)$ equation, where $p(r)$ and $\epsilon(r)$ are replaced by $P_{\mathrm{tot}}(r)$ and $\epsilon_{\mathrm{tot}}(r)$, respectively. The resulting surface value $y_R$ and stellar compactness $C$ are subsequently substituted into Eq.~\eqref{eq:k2} to compute  $k_2$ and  $\Lambda$. {It is worth noting that, in the two--fluid formalism, the SQM and DM components may have different surface radii. Therefore, the radius used to evaluate $y_R$, the compactness $C$, the Love number $k_2$, and consequently the tidal deformability $\Lambda$ is the outermost radius of the configuration, given by $R=\max(R_{\rm SQM},R_{\rm DM}).$
Thus, the relevant stellar radius may correspond either to the SQM radius or to the DM radius, depending on which component extends farther outward.}

Figure~\ref{tidal-twofluid} shows $\Lambda$ as a function of $M/M_\odot$ for various values of $f_r$ and $m_D$. Similar to the one--fluid case, we compare our curves with the GW170817 gray area ($70 \le \Lambda_{1.4 M_\odot} \le 580$) \cite{Abbott2018} and the prospective pink band ($120 \le \Lambda_{1.4 M_\odot} \le 400$) \cite{Annala2018}. To clarify this point, an important distinction must be made regarding the nature of these limits. The GW170817 boundary is a direct observational constraint on the star's overall gravitational tidal response. Although the standard LIGO/Virgo analysis does not explicitly assume an internal DM component, the measured $\Lambda$ is a global, model--independent parameter determined solely by the total mass--energy distribution and its external metric perturbation. Thus, the aggregate tidal deformability of our gravitationally coupled two--fluid system must physically satisfy this observational limit. On the other hand, the tighter pink band was originally proposed using traditional one--fluid models. We adopt this limit not as a rigid boundary, but as a demanding qualitative test for our multi--fluid configurations. Specifically, we investigate whether our model can simultaneously meet this prospective $\Lambda$ constraint alongside strict causality and  $2.0\,M_\odot$ mass constraint. The plots show that even without direct interactions, DM still produces a gravitational pull that makes the star more compact. Consequently, this gravitational effect shifts the $\Lambda-M$ curves downward. 
The numerical results reveal that the behavior of our two--fluid model is highly sensitive to both the values of $m_D$ and  $f_r$. Specifically, as the value of $f_r$ increases, the range of $m_D$ that can satisfy our constraints shifts systematically (Table \ref{tab:comparison_tidal2}). 
For a small value of $f_r = 2\%$, the pink band is populated by relatively light dark particles, roughly in the range of $m_D \approx 210 - 230\text{ MeV}$. In this panel, a lighter particle (like $m_D = 200\text{ MeV}$) yields a tidal deformability $\Lambda_{1.4 M_\odot}$ that is too high, while a heavier particle (like $m_D = 240\text{ MeV}$) pushes the curves below the lower boundary of the pink band.
As we increase $f_r$, we must use heavier dark particles to stay within the same boundaries. At $f_r = 4\%$, the allowed range for $m_D$ shifts to approximately $260 - 310\text{ MeV}$. This upward trend continues to $m_D \approx 310 - 350\text{ MeV}$ for $f_r = 6\%$, and finally reaches $m_D \approx 330 - 380\text{ MeV}$ when $f_r = 8\%$. These systematic shifts show a clear degeneracy between $f_r$ and   $m_D$. This highlights a major advantage of our non--interacting two--fluid model. Removing the direct interaction reduces the free parameters, making the model simpler and more robust. Our results show that gravitational coupling alone is enough to increase the compactness of the star. Even without complex forces between the two fluids, this simple setup easily satisfies both current GW170817 constraints and future tighter limits. This high adaptability makes two--fluid framework a good candidate for interpreting future high--precision gravitational wave data, especially when searching for DM signatures in compact stars. {Additionally, while a weak non--gravitational interaction may exist between quarks and scalar DM, our results suggest that it would likely be sufficiently weak for the resulting stellar configuration to remain close to the effectively decoupled two--fluid behavior rather than approaching a one--fluid regime. This interpretation is supported by the better compatibility of the two--fluid configurations with the current and anticipated constraints on $\Lambda$.
	}

\section{Conclusion}
\label{sec:conclusion}
In this work, we have performed a systematic and comparative study of strange quark stars (SQSs) admixed with scalar dark matter (DM). {Specifically, we have investigated whether the model with quark--DM interactions or the non--interacting one is more compatible with current as well as future expected observational results from gravitational waves and pulsars.} To this end, we have contrasted two fundamentally distinct modeling frameworks: an interacting one--fluid model and a non--interacting two--fluid model. The former assumes a fully mixed state where strange quark matter (SQM) and DM are thermodynamically coupled via a microscopic Yukawa--type interactions and Bose--Einstein condensation (BEC) pressure. The latter treats the SQM and DM as independent fluid components that interact solely through the shared gravitational field. For both frameworks, we have computed the stellar structure parameters, mass--radius relations, and dimensionless tidal deformability ($\Lambda$) to evaluate their viability against current astrophysical observations and anticipated limits from next--generation gravitational wave detectors.

{To present our analysis systematically, we have structured our investigation in a step--by--step manner. First, we have outlined the thermodynamic formulations and equations of state (EOS) for both the interacting one--fluid and non--interacting two--fluid models. In this initial stage, we have shown that the requirement of causality imposes strict limits on the two--fluid model. Specifically, we have demonstrated that to keep the speed of sound physically acceptable under DM pressures, the scalar DM particle mass ($m_D$) cannot be arbitrarily small and must have a clear lower limit.
Following this analysis and setting the gap parameter $\Delta=100 \,\text{ MeV}$ (to achieve massive pulsars), we have addressed the primary motivation of our study by evaluating  $\Lambda$. While both models have been shown to comfortably meet the current observational constraint of $\Lambda_{1.4 M_\odot} \lesssim 580$ from the GW170817 event \cite{Abbott2018}, we have discussed how this threshold is expected to become significantly tighter ($\Lambda_{1.4 M_\odot} \lesssim 400$) with the advent of next--generation gravitational wave detectors and advanced X--ray missions \cite{Annala2018}. Crucially, our comparative analysis has revealed that the interacting one--fluid model struggles to satisfy this anticipated tighter boundary unless confined to highly restricted parameter choices. Conversely, the non--interacting two--fluid model satisfies these prospective limits across a much broader configuration space, suggesting that purely gravitational coupling is more compatible with future $\Lambda$ bounds. To further support this conclusion, we show how varying the model parameters affects our results.} 

\begin{figure*}[htbp]
	\centering
	\includegraphics[width=17cm]{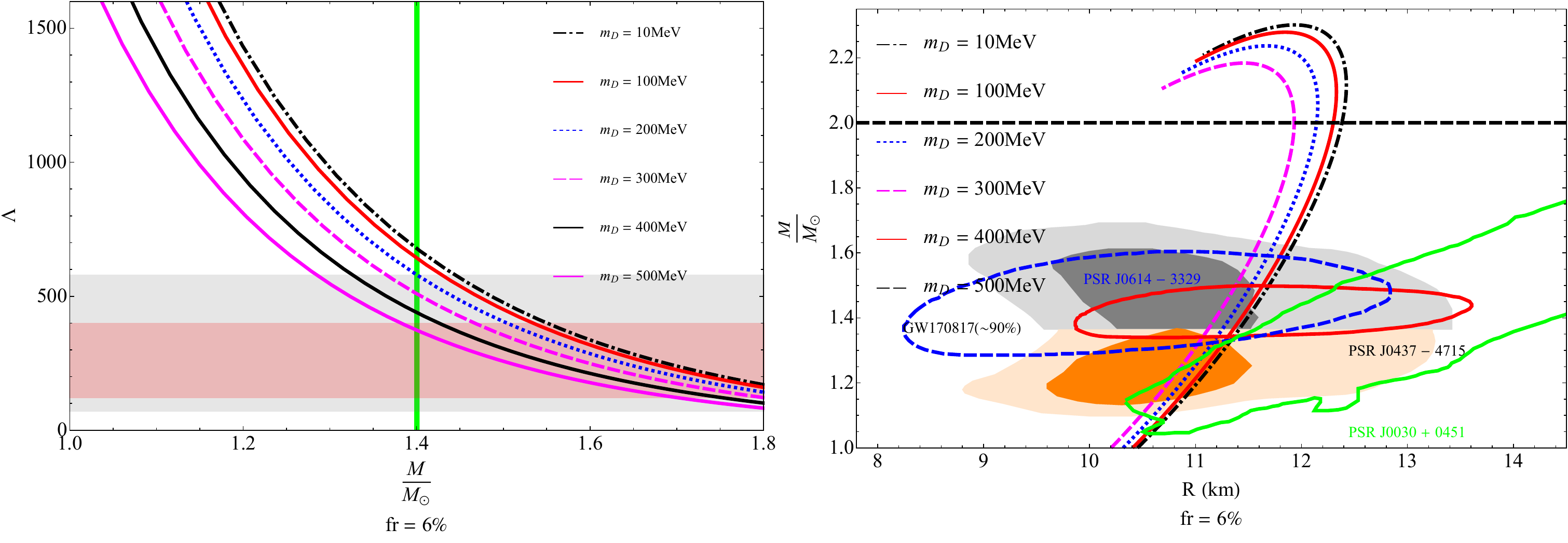}
	\caption{{$\Lambda-M$ (left panel) and $M-R$   (right panel) diagrams for $fr=6\%$, $\Delta=110\,\text{MeV}$ and different values of $m_D$ in the interacting formalism. The horizontal line in $M-R$ plot shows the value $M=2.0\,M_\odot$.}}
	\label{mrtidal110onefluid}
\end{figure*}

\begin{figure*}[htbp]
	\centering
	\includegraphics[width=17cm]{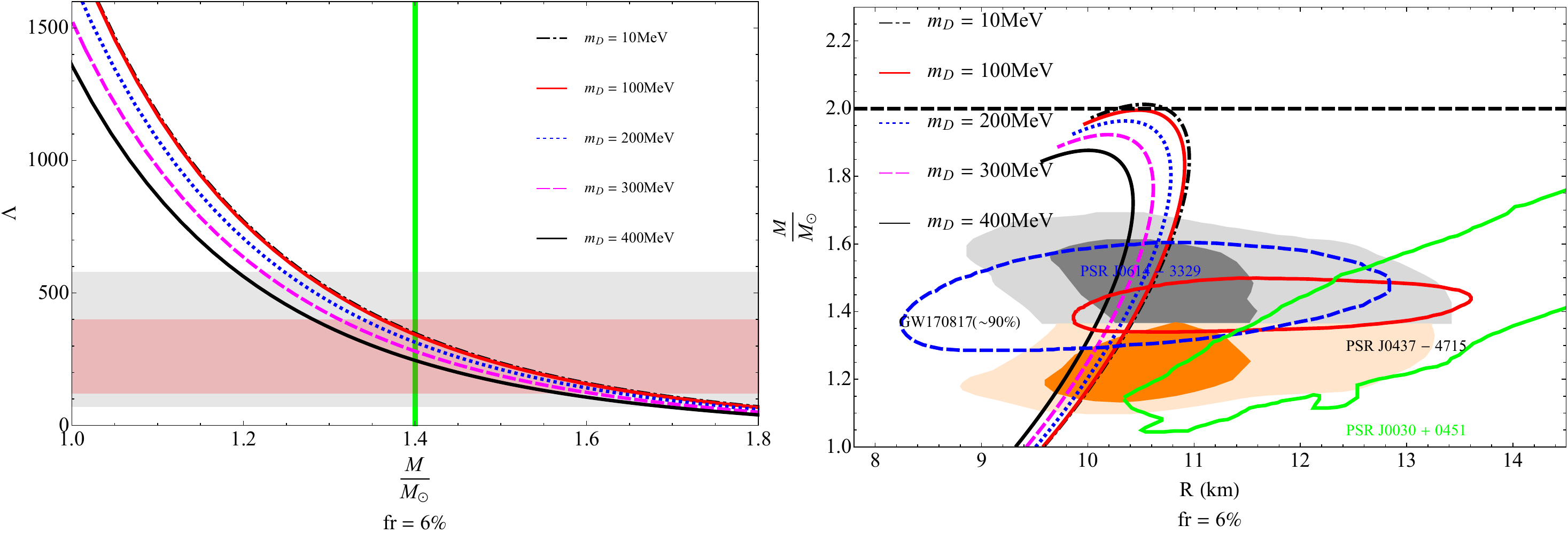}
	\caption{{$\Lambda-M$ (left panel) and $M-R$   (right panel) diagrams for $fr=6\%$, $\Delta=90\,\text{MeV}$ and different values of $m_D$ in the interacting formalism. The horizontal line in $M-R$ plot shows the value $M=2.0\,M_\odot$.}}
	\label{mrtidal90onefluid}
\end{figure*}

{In the interacting one-fluid framework, any modification of the parameter space essentially leads to one of two outcomes:
\begin{itemize}
	\item It makes the EOS stiffer, such as increasing the gap parameter from $\Delta = 100$ to $110\,\text{MeV}$. While this increases the maximum mass, it also raises the tidal deformability, moving it further away from the predicted tighter constraint $\Lambda_{1.4 M_\odot} \lesssim 400$ (Fig.~\ref{mrtidal110onefluid}).
	\item It softens the EOS, such as reducing the gap parameter from $\Delta = 100$ to $ 90\,\text{MeV}$. Although this satisfies the tighter $\Lambda$ limit, the configurations fail to reach  $M_{\text{TOV}}>2.0\,M_\odot$ (Fig.~\ref{mrtidal90onefluid}).
\end{itemize}}

\begin{figure*}[htbp]
	\centering
	\includegraphics[width=17cm]{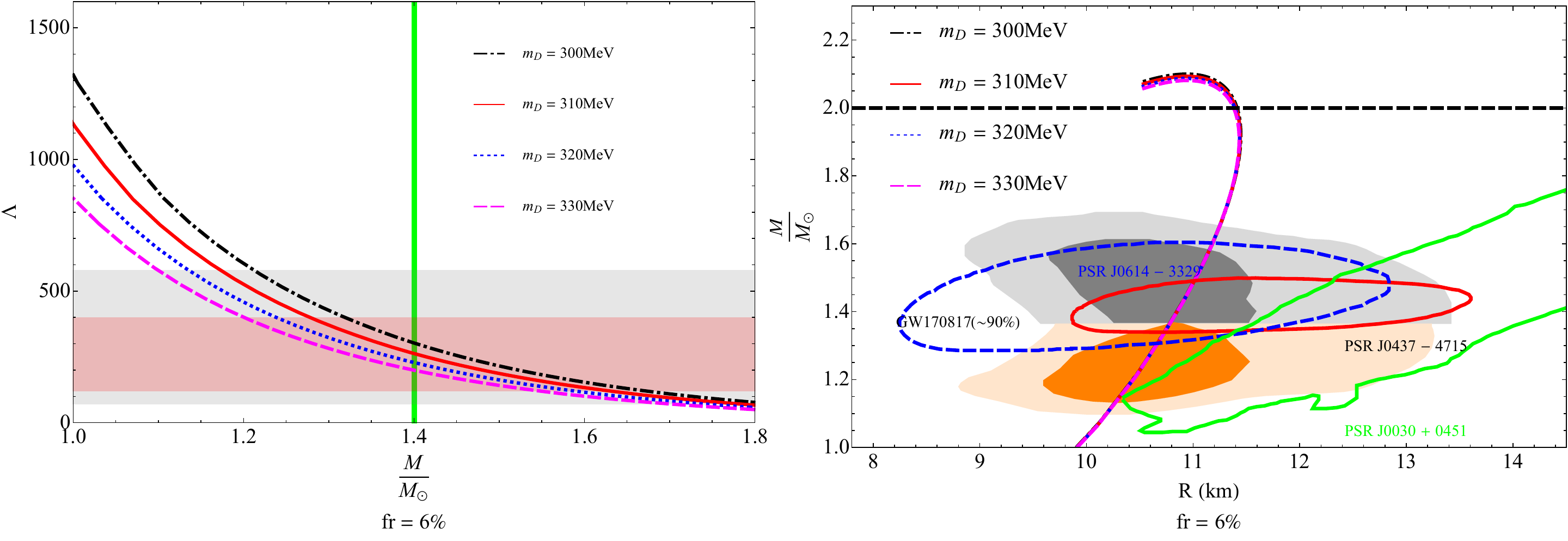}
	\caption{{$\Lambda-M$ (left panel) and $M-R$   (right panel) diagrams for $fr=6\%$, $\Delta=110\,\text{MeV}$ and different values of $m_D$ in the non--interacting formalism. The horizontal line in $M-R$ plot shows the value $M=2.0\,M_\odot$.}}
	\label{mrtidal110twofluid}
\end{figure*}

{In contrast, the non-interacting two-fluid model naturally overcomes this challenge. As shown in Fig.~\ref{mrtidal110twofluid}, stiffening the quark EOS with $\Delta = 110\,\text{MeV}$ allows the star to simultaneously achieve $M_{\rm TOV} > 2.0\,M_\odot$ and satisfy the tighter $\Lambda_{1.4 M_\odot} \lesssim 400$ threshold. It should be mentioned that while other parameter variations could be considered, we limit our presentation to these representative cases for brevity, as our focus is to compare the general physical behavior of the two frameworks rather than to constrain the entire parameter space.}

{Ultimately, our results show that if scalar DM exists, the non-interacting model provides a favorable framework that remains compatible with both pulsar measurements and anticipated tighter constraints from future gravitational-wave observations.}

\newpage
\section*{Acknowledgements}
We wish to thank Shiraz University Research
Council. This work is based upon research funded by Iran National
Science Foundation (INSF) under project No. 4044248.

\end{document}